# Binder chemistry sets the interfacial balance constant in $CsPbBr_3$ nanocrystal supercapacitor electrodes

*Arun Kumar*[†], *Monojit Bag*[†, ‡, *].

†Advanced Research in Electrochemical Impedance Spectroscopy Laboratory, Indian Institute of Technology Roorkee, Roorkee 247667, India

‡Centre for Nanotechnology, Indian Institute of Technology Roorkee, Roorkee 247667, India

^Department of Physics, Banaras Hindu University, Varanasi 221005, India

## Abstract

Polymer binders in composite supercapacitor electrodes are normally selected by convention and treated as inert structural components. Previous work on lead-free tin halide perovskites showed that the binder instead sets the electrolyte concentration at which capacitance is maximised, following the relationship $100 \times [Li^{+}]_{opt} + PVDF_{wt\%} = \xi_{Int}$ with $\xi_{Int}$ = 25 ± 2.5, established by varying the loading of a single polymer. Whether $\xi_{Int}$ is universal or specific to that polymer has not been tested. Here, four binders spanning fluorinated (PVDF), carboxylic (PAA), cellulosic (CMC) and sulfonic (PEDOT:PSS) chemistry are compared on $CsPbBr_3$ nanocrystal electrodes at four LiTFSI concentrations in acetonitrile, with binder loading fixed at 15 wt% and all other formulation variables held constant. The relationship applies to $CsPbBr_3$: PVDF at 15 wt% gives an optimum at 0.10–0.15 M and 112 F $g^{-1}$, against 126 F $g^{-1}$ reported for $CsSnCl_3$ under the same conditions, extending the result to a different B-site cation, halide and crystal system. The constant is not the same for all binders. PVDF and CMC optimise at 0.10 M, giving $\xi_{Int}$ = 25, while PAA and PEDOT:PSS optimise at 0.15 M, giving $\xi_{Int}$ = 30, with maximum values of 188 F $g^{-1}$ for PAA and 146 mF $cm^{-2}$ for PEDOT:PSS. The two binders showing the shift carry ionisable acid groups at high density, indicating that $\xi_{Int}$ expressed in weight percent requires a binder-specific value. Charge transfer resistance falls monotonically from 75–180 Ω at 0.05 M to 19–29 Ω at 0.20 M and is lowest where capacitance is poorest, so the decline above the optimum is not caused by restricted ion supply. Structural and spectroscopic measurements show that the binder also determines the chemical state of the perovskite before any potential is applied: surface Cs/Pb varies by a factor of 2.8 across the four electrodes, metallic lead is detected in the CMC and PEDOT:PSS electrodes. In all four electrodes $CsPbBr_3$ converts to $PbBr_2$ and CsBr during electrochemical characterisation, and the surface lead content of the PEDOT:PSS electrode decreases. The interfacial optimum and the chemical stability of the perovskite are therefore independent properties, and binder chemistry rather than binder mass fraction alone determines where the electrolyte optimum lies.

## 1. Introduction

Electrochemical capacitors and batteries occupy separate regions of the energy–power space. Batteries store 100–250 Wh $kg^{-1}$ but deliver power in the range of $10^2$–$10^3$ W $kg^{-1}$, while electrochemical capacitors deliver $10^3$–$10^4$ W $kg^{-1}$ at energy densities below about 10 Wh $kg^{-1}$.[1] The difference arises from the charge storage mechanism. In an electrical double-layer capacitor, ions accumulate electrostatically at the electrode surface, so the process is fast and reversible over $10^5$–$10^6$ cycles but limited by the accessible surface area.[2] Raising the stored energy without losing the rate capability requires charge storage that extends beyond the surface layer, which is the reason materials showing reversible faradaic reactions at or near the electrode surface have been examined as capacitor electrodes.[3] Metal halide perovskites are one such class. In these materials, both electronic and ionic transport occur at room temperature, and the ionic component is unusually fast. Halide vacancy migration in lead halide perovskites proceeds with activation energies reported between 0.1 and 0.6 eV depending on composition and measurement methode,[4,5] and ionic conductivities of $10^{-9}$ to $10^{-7}$ S $cm^{-1}$ have been measured under bias.[6] Charge therefore redistributes within the lattice on the timescale of a charge–discharge cycle rather than being confined to the outer surface. In the all-inorganic cesium lead halides, this behaviour is combined with better thermal tolerance than the hybrid organic–inorganic analogues, since no volatile organic cation is present.[7] $CsPbBr_3$ in particular can be prepared as monodisperse nanocrystals by ligand-assisted reprecipitation at room temperature, giving surface areas well above those of bulk-processed powders.[8] Halide perovskite electrodes have been reported with specific capacitances spanning roughly 40 to 300 F $g^{-1}$.[9–12] The width of that range reflects differences in electrolyte identity and concentration, electrode formulation, potential window and current density rather than differences in the perovskite alone, which makes direct comparison between reports difficult. Photo-rechargeable and integrated photo-supercapacitor architectures based on the same materials have also been described, extending the application space beyond conventional energy storage.[13] The property that makes these materials attractive also limits them. Under an applied field, mobile halide ions accumulate at grain boundaries, interfaces and electrode contacts, and the resulting compositional gradients drive irreversible structural change.[14,15] In mixed-halide compositions, field-driven separation of the halide sublattice has been observed directly during electrochemical cycling.[16] In the single-halide bromide, the accepted decomposition route is loss of the perovskite framework to its binary constituents, $CsPbBr_3 \rightarrow CsBr + PbBr_2$, with $PbBr_2$ appearing as a crystalline product in diffraction.[16] Because a supercapacitor electrode is held under bias for far longer, and cycled far more often, than a photovoltaic absorber, the electrochemical environment is a demanding one for these materials. Most work addressing this problem has focused on the perovskite: compositional tuning, A-site or B-site substitution, surface passivation with modified ligands, or encapsulation.[17,18] The electrode formulation itself has received less scrutiny. A composite electrode is not the active material alone. It contains 10–20 wt% polymer binder and a comparable fraction of conductive carbon, and the binder is normally selected by convention: poly(vinylidene fluoride) for organic electrolytes, carboxymethyl cellulose or poly(acrylic acid) for aqueous systems, and conducting polymers where electronic percolation is limiting.[19–23] For graphite or transition metal oxide electrodes this treatment is reasonable, because the binder is largely chemically inert toward the active material and its role is mechanical. A halide perovskite surface does not permit that assumption. The nanocrystal surface terminates in undercoordinated $Pb^{2+}$ sites and halide vacancies, and these sites coordinate readily to Lewis-basic functional groups.[24] Carboxylate, hydroxyl and sulfonate groups are all capable of binding to surface lead, and the same coordination chemistry underlies the ligand systems used to passivate perovskite nanocrystals during synthesis.[25] A binder carrying these groups is therefore a chemical participant at the electrode interface, not a spectator. It can passivate surface defects, it can extract halide, and where the polymer is a strong acid it can attack the lattice directly.

Recent work on lead-free tin halide perovskites established that this coupling follows a quantitative relationship.[26] Across a matrix of PVDF loadings from 5 to 20 wt% and LiTFSI concentrations from 0.05 to 0.20 M in $CsSnCl_3$ electrodes, the concentration at which capacitance is maximised decreases linearly with increasing binder content, and every optimum satisfies

$$100 \times [\mathrm{Li}^+]_{opt} + \mathrm{binder}_{\mathrm{wt\%}} = \xi_{Int} \qquad (1)$$

with $\xi_{Int}$ = 25 ± 2.5, the uncertainty following from the 0.05 M resolution of the concentration grid. In normalised form, with $\lambda = [Li^+]_{opt}/0.25$ M and $\theta = binder_{wt\%}/25$, this becomes $\lambda + \theta = 1$. The same constant was obtained for the hybrid analogue $MASnCl_3$, indicating that the optimum is set at the polymer–electrolyte interface rather than by the A-site chemistry of the lattice. Atomistic simulations attributed the relationship to a finite two-dimensional accommodation capacity at the interface: the adsorbed polymer compresses the site-to-site variation in $Li^+$ adsorption energy from 2.63 to 0.19 eV and confines lithium beneath a fluorine-rich overlayer, while occupying interfacial area that would otherwise accommodate lithium. Beyond an optimum coverage, the loss of available area outweighs the improvement in coordination. Equation (1) was obtained by varying the loading of a single polymer. Whether $\xi_{Int}$ is a universal constant or a property of the specific binder has not been tested, and the distinction matters because the relationship is expressed in weight percent. Equal mass fractions of different polymers deliver very different numbers of functional groups to the interface, so a constant defined on mass need not transfer between binder chemistries. At the 15 wt% loading used throughout the present work, equation (1) with $\xi_{Int}$ = 25 predicts a capacitance optimum at 0.10 M LiTFSI for every binder. This work addresses that question. Four binders spanning four functional-group families are compared on $CsPbBr_3$ nanocrystal electrodes: PVDF ($-CF_2-$, no ionisable groups), poly(acrylic acid) (–COOH), carboxymethyl cellulose ($-COO^-$ and –OH on a cellulose backbone), and PEDOT:PSS ($-SO_3H$ on a conducting polymer). Each is examined at four LiTFSI concentrations in acetonitrile — 0.05, 0.10, 0.15 and 0.20 M — giving sixteen electrode–electrolyte combinations. Binder loading, active material loading, conductive carbon content, solvent and casting procedure are held constant across the series, so that binder chemistry and electrolyte concentration are the only variables. $CsPbBr_3$ was selected rather than a lead-free analogue so that binder effects could be measured without the competing $Sn^{2+} \rightarrow Sn^{4+}$ oxidation that limits tin-based perovskites under ambient and electrochemical conditions.[27] Three results are reported. First, the interfacial balance relationship holds on $CsPbBr_3$. With PVDF at 15 wt%, the electrode formulation used in the earlier study, the capacitance optimum falls at 0.10–0.15 M and reaches 112 F $g^{-1}$, against 126 F $g^{-1}$ measured for $CsSnCl_3$ under the same formulation and concentration.[26] Since $CsPbBr_3$ differs from the earlier systems in B-site cation, halide and crystal system, the relationship is not specific to the tin chloride lattice. Second, $\xi_{Int}$ is binder-dependent. PVDF and CMC optimise at 0.10 M, consistent with $\xi_{Int}$ = 25, while PAA and PEDOT:PSS optimise one grid step higher at 0.15 M, corresponding to $\xi_{Int}$ = 30, with the highest specific capacitance in the series of 188 F $g^{-1}$ for PAA and the highest areal capacitance of 146 mF $cm^{-2}$ for PEDOT:PSS. The two binders showing the shift are those carrying ionisable acid groups. Third, charge transfer resistance decreases monotonically with LiTFSI concentration, from 75–180 Ω at 0.05 M to 19–29 Ω at 0.20 M, and is lowest where capacitance is poorest, so the decline above the optimum is not caused by limited ion supply. Structural and spectroscopic measurements show that the binder also determines the chemical state of the nanocrystal surface before any potential is applied, and that active material is lost from all four electrodes during electrochemical operation.

## 2. Experimental

### 2.1 Materials

Cesium bromide (99.6%, TCI Chemicals), lead(II) bromide (99.9%, Sigma Aldrich), oleic acid (99%, SRL chemicals), oleylamine (99%, SRL chemicals), dimethylformamide (DMF, anhydrous 99.8%, Sigma Aldrich), poly(vinylidene fluoride) (Sigma Aldrich), poly(acrylic acid) (Sigma Aldrich), sodium carboxymethyl cellulose (TCI Chemicals),[28] PEDOT:PSS 3.0–4.0 wt% dispersion in $H_2O$ (Sigma Aldrich), Super P conductive carbon (Sigma Aldrich), *N*-methyl-2-pyrrolidone (NMP, 99%, SRL Chemicals), Lithium bis(trifluoromethane sulfonic)imide (99%, Sigma Aldrich) and anhydrous acetonitrile (99%, SRL Chemicals) were used as received.

### 2.2 Synthesis of $CsPbBr_3$ nanocrystals

$CsPbBr_3$ nanocrystals were prepared by ligand-assisted reprecipitation.[29] CsBr and $PbBr_2$ were combined in a 1:1 molar ratio, 0.4 mmol of each, and dissolved in 10 mL of anhydrous DMF, giving 0.04 M in each salt. Oleylamine and oleic acid were added in a 1:2 volume ratio. The mixture was stirred at 1000 rpm and 50 °C for 15 min and then sonicated for 15 min, and the two steps were repeated until the solution was clear. A 100 µL aliquot of the precursor was added dropwise to 1 mL of anhydrous toluene at room temperature under vigorous stirring, producing immediate nucleation. The suspension was centrifuged at 10 000 rpm for 10 min. The recovered powder was dried under vacuum at 70 °C for 12 h.

### 2.3 Electrode preparation

Electrodes were prepared with a fixed composition of 70:15:15 by weight of $CsPbBr_3$ nanocrystals, Super P carbon and binder. Each component was combined in *N*-methyl-2-pyrrolidone and stirred for 24 h at room temperature to form a slurry. The slurry was applied by hand brushing onto 1 cm × 1 cm graphite substrates and dried at 80 °C. Active material loading was approximately 1 mg per electrode, giving a geometric loading near 1 mg $cm^{-2}$.[30] The same solvent was used for all four binders so that solvent identity would not vary across the series. The four polymers do not behave identically in NMP. PVDF and PAA dissolve to give homogeneous solutions. CMC and PEDOT:PSS do not dissolve in NMP and form dispersions, so in these two systems the binder is present as dispersed material rather than as a continuous polymer film. This difference in binder distribution state is a co-variable that could not be eliminated without changing the solvent, and it is considered alongside functional-group effects in the discussion. PEDOT:PSS was supplied as a 3.0–4.0 wt% aqueous dispersion and therefore introduced water into the slurry. At a mean solids content of 3.5 wt%, delivering the 0.21 mg of binder required for a nominal 1.4 mg electrode requires approximately 6 mg of dispersion, of which approximately 5.8 mg is water, exceeding the total solid mass of the electrode by a factor of four. No comparable water addition occurs for the other three binders. **Figure 1** shows the synthesis and electrode fabrication sequence schematically.

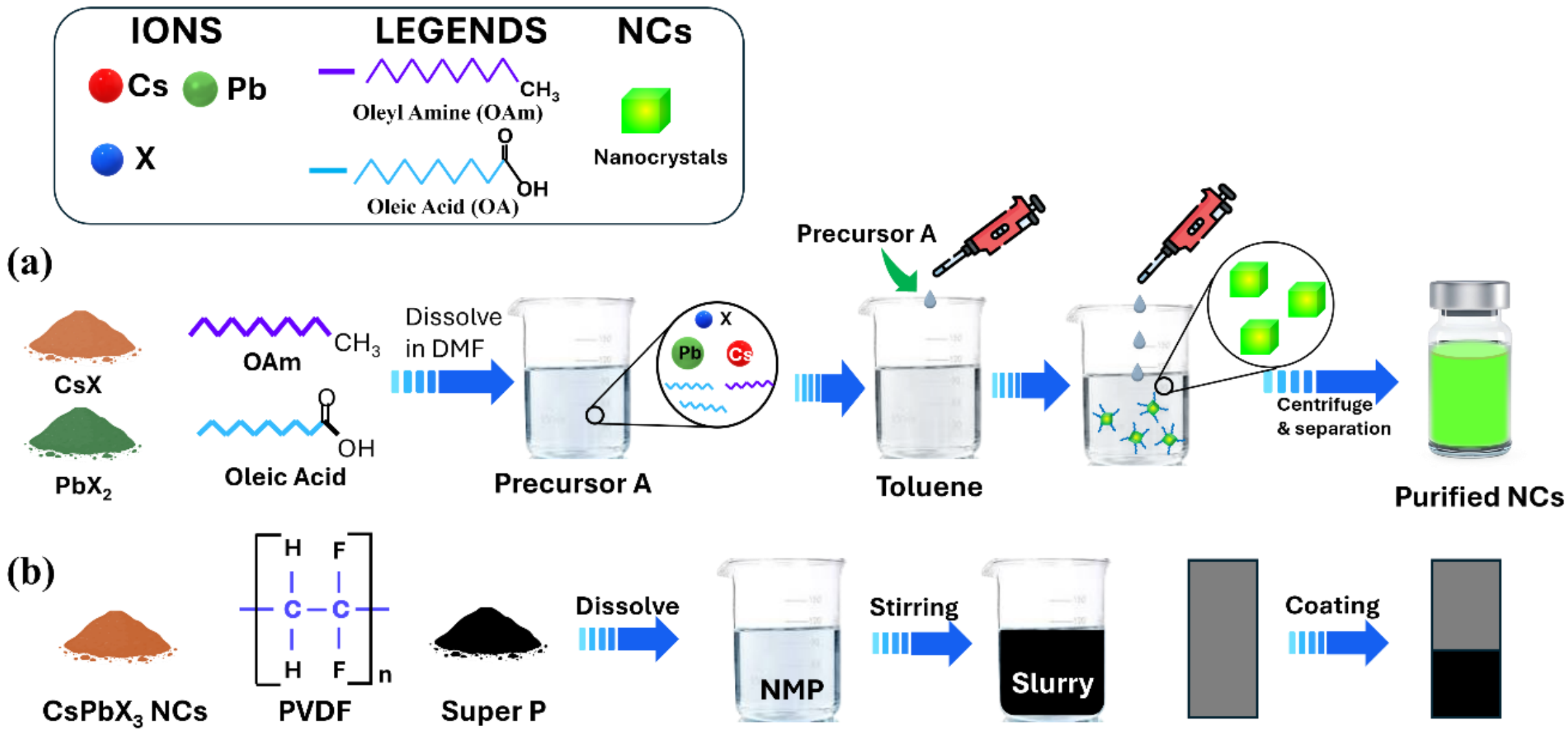

**Figure 1.** Schematic of (a) ligand-assisted reprecipitation synthesis of $CsPbBr_3$ nanocrystals and (b) electrode fabrication, in which the nanocrystals, Super P conductive carbon and polymer binder are combined in NMP at 70:15:15 by weight and applied to graphite substrates.

### 2.4 Structural and morphological characterisation

X-ray diffraction patterns were recorded on a Bruker D8 Advance diffractometer with Cu Kα radiation (λ = 1.5406 Å). Patterns were indexed to orthorhombic $CsPbBr_3$ in the Pnma setting with lattice parameters a = 8.244 Å, b = 11.735 Å and c = 8.198 Å [Ref: COD 4510745]. Transmission electron microscopy, high-resolution imaging and selected-area electron diffraction were performed on a JEOL JEM-3200FS. Nanocrystals were dispersed in toluene and drop-cast onto carbon-coated copper grids. Surface morphology of the electrodes and elemental mapping were obtained on a Carl Zeiss Gemini field-emission scanning electron microscope with an attached energy-dispersive X-ray spectrometer. X-ray photoelectron spectra were recorded on a PHI VersaProbe III spectrometer with monochromatic Al Kα radiation (1486.6 eV).

### 2.5 Electrochemical measurements

All measurements were made in a three-electrode cell containing lithium bis(trifluoromethanesulfonyl)imide (LiTFSI) in anhydrous acetonitrile at concentrations of 0.05, 0.10, 0.15 and 0.20 M, with the coated graphite as working electrode, Ag/AgCl as reference and platinum as counter electrode, on an Autolab potentiostat. Cyclic voltammograms were recorded at scan rates from 5 to 140 mV $s^{-1}$. Galvanostatic charge–discharge curves were recorded at current densities from 0.2 to 0.7 A $g^{-1}$. Impedance spectra were recorded from 0.01 Hz to 100 kHz, at open-circuit potential and at DC bias values of 0, 0.2, 0.4, 0.6, 0.8 and 1.0 V against Ag/AgCl. Specific capacitance, areal capacitance, energy density and power density were calculated using equations S1–S6, and ionic conductivity and diffusion coefficients from the impedance data using the Bandara–Mellander treatment given in equations S7–S12.[31]

## 3. Results and Discussion

### 3.1 Structure and morphology of the nanocrystals and pristine electrodes

The diffraction pattern of the as-synthesised powder is shown in the lowest trace of Figure 2a. Reflections are indexed to orthorhombic $CsPbBr_3$ in the *Pnma* setting with lattice parameters a = 8.244 Å, b = 11.735 Å and c = 8.198 Å (CIF 4510745).[32] The observed positions correspond

to (020) at 15.09°, (121) at 21.49°, (122) at 28.66°, (040) at 30.45°, (202) at 30.73°, (222) at 34.40°, (240) at 37.60°, (241) at 39.24°, (242) at 43.80° and (060) at 46.40°. Resolution of the (040) and (202) reflections as a doublet near 30.5° establishes the orthorhombic phase, since these two reflections are degenerate in the cubic setting and their separation reflects the difference between the b and the a, c axes. The reflections are broad, consistent with nanoscale crystalline domains. A weak reflection at 12.8° is marked with an asterisk in Figure 2a and is assigned to $Cs_4PbBr_6$, a 0D cesium-rich phase that forms alongside $CsPbBr_3$ when the Cs:Pb ratio in the precursor solution exceeds unity.[33] The as-synthesised powder therefore contains a minor secondary phase. This reflection is not detected in any of the four electrode patterns.

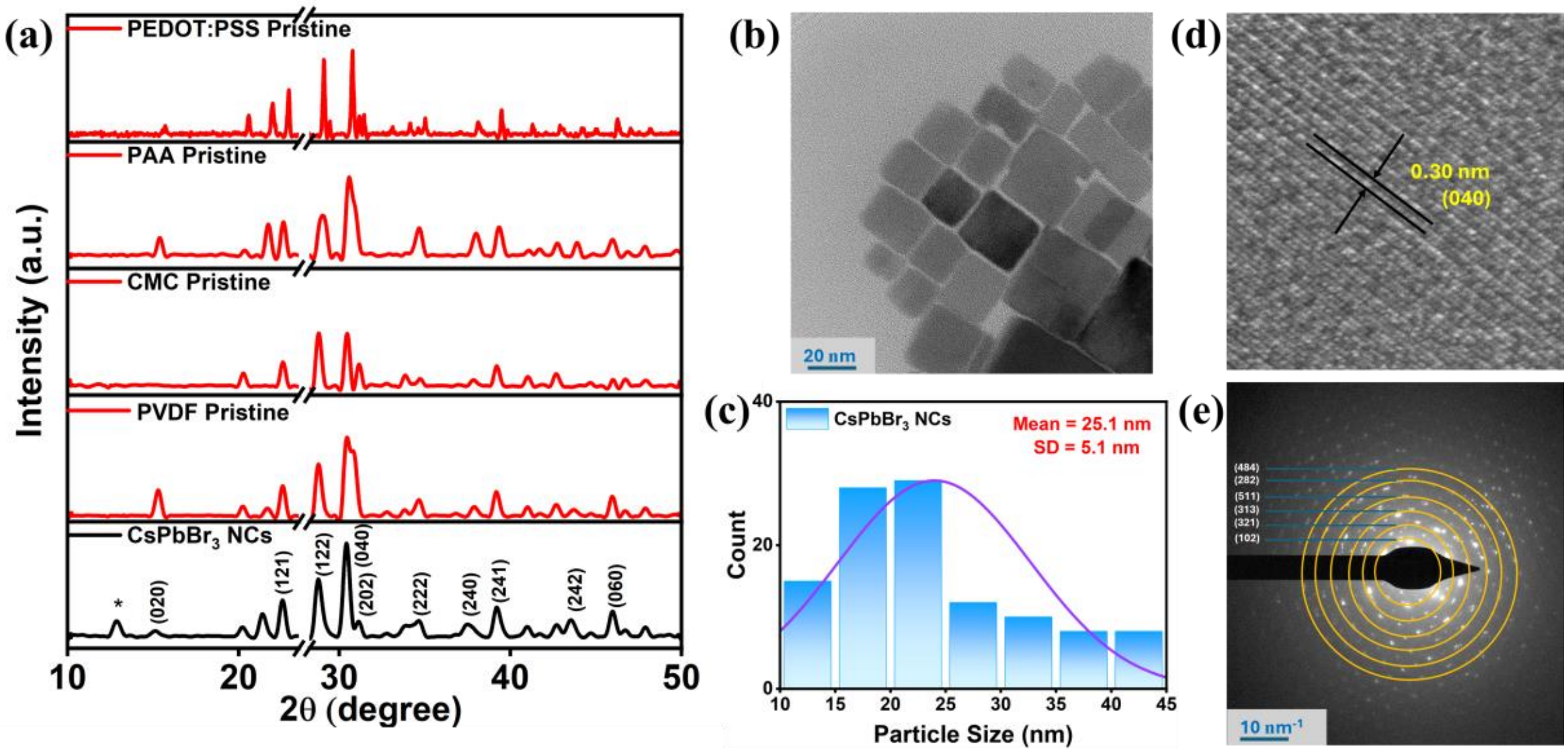

**Figure 2.** Structural and morphological characterisation of $CsPbBr_3$ nanocrystals (NCs) and pristine electrodes. (a) X-ray diffraction patterns of the as-synthesised NCs (black) and of electrodes prepared with PVDF, CMC, PAA and PEDOT:PSS (red). Reflections are indexed to orthorhombic $CsPbBr_3$ (*Pnma*, a = 8.244 Å, b = 11.735 Å, c = 8.198 Å); the asterisk marks a $Cs_4PbBr_6$ reflection at 12.8° present in the powder. (b) Bright-field TEM image of the nanocrystals. (c) Particle size distribution measured over 110 particles, giving a mean edge length of 25.1 nm with a standard deviation of 5.1 nm. (d) Selected-area electron diffraction pattern, with measured d-spacings of 3.637, 2.382, 1.905, 1.597, 1.305 and 1.042 Å indexed to the (102), (321), (313), (511), (282) and (484) reflections. (e) High-resolution image showing lattice fringes with a spacing of 0.30 nm, corresponding to the (040) planes.

Transmission electron microscopy in Figure 2b shows particles with square and rectangular projections and well-defined edges, consistent with the cuboidal habit reported for $CsPbBr_3$ nanocrystals prepared by ligand-assisted reprecipitation.[34] The size distribution in Figure 2c, measured over 110 particles, gives a mean edge length of 25.1 nm with a standard deviation of 5.1 nm, corresponding to a relative width of 20%. The distribution is asymmetric, with a tail extending to 45 nm. The selected-area electron diffraction pattern in Figure 2d consists of discrete spots arranged on concentric rings rather than continuous rings, which indicates a limited number of randomly oriented crystals within the selected area, as expected for particles of this size. Measured d-spacings of 3.637, 2.382, 1.905, 1.597, 1.305 and 1.042 Å are indexed to the (102), (321), (313), (511), (282) and (484) reflections of the same orthorhombic cell, in agreement with the diffraction assignment. The high-resolution image in Figure 2e shows continuous lattice fringes across the imaged region, indicating that individual particles are single crystalline rather than polycrystalline aggregates. The measured fringe spacing of 0.30 nm corresponds to the (040) planes, for which the calculated interplanar spacing is 0.293 nm. The four pristine electrode patterns are shown as the upper traces of Figure 2a. All four electrodes retain the full set of $CsPbBr_3$ reflections, with peak positions unchanged from the powder. The reflections are considerably narrower than those of the powder, *with the (121) reflection full width at half maximum decreasing from 0.51 to 0.46,* which indicates that the crystalline domain size increased during slurry preparation and drying. NMP is a coordinating solvent capable of dissolving $PbBr_2$ and displacing the oleate and oleylammonium surface ligands,[25,35] and the loss of ligand coverage permits Ostwald ripening of the nanocrystals during the stirring and drying steps. The nanocrystals in the finished electrode are therefore larger than those characterized in Figure 2b–e, and the specific surface area available for charge storage is correspondingly lower than the as-synthesised dimensions would suggest. Field-emission scanning electron micrographs of the pristine electrodes are shown in Figure 6e and Figure 6g and in Figure S9. The surfaces consist of faceted crystals with dimensions between approximately 200 nm and 1 μm distributed over a finer particulate matrix of conductive carbon and binder. The dimensions of these crystals exceed the 25.1 nm mean measured for the as-synthesised nanocrystals by more than an order of magnitude, in agreement with the peak narrowing described above. Coarsening during electrode fabrication is therefore observed by two independent methods.

### 3.2 Electrochemical performance

Galvanostatic charge–discharge profiles at 0.2 A g$^{-1}$ are shown in Figure 3a–d and cyclic voltammograms recorded at 5 mV s$^{-1}$ in Figure 3e–h, with all four LiTFSI concentrations overlaid in each panel. Profiles measured at 0.3–0.7 A g$^{-1}$ and at scan rates from 10 to 140 mV s$^{-1}$ are collected in Figure S1-S4 and Figure S5-S8.

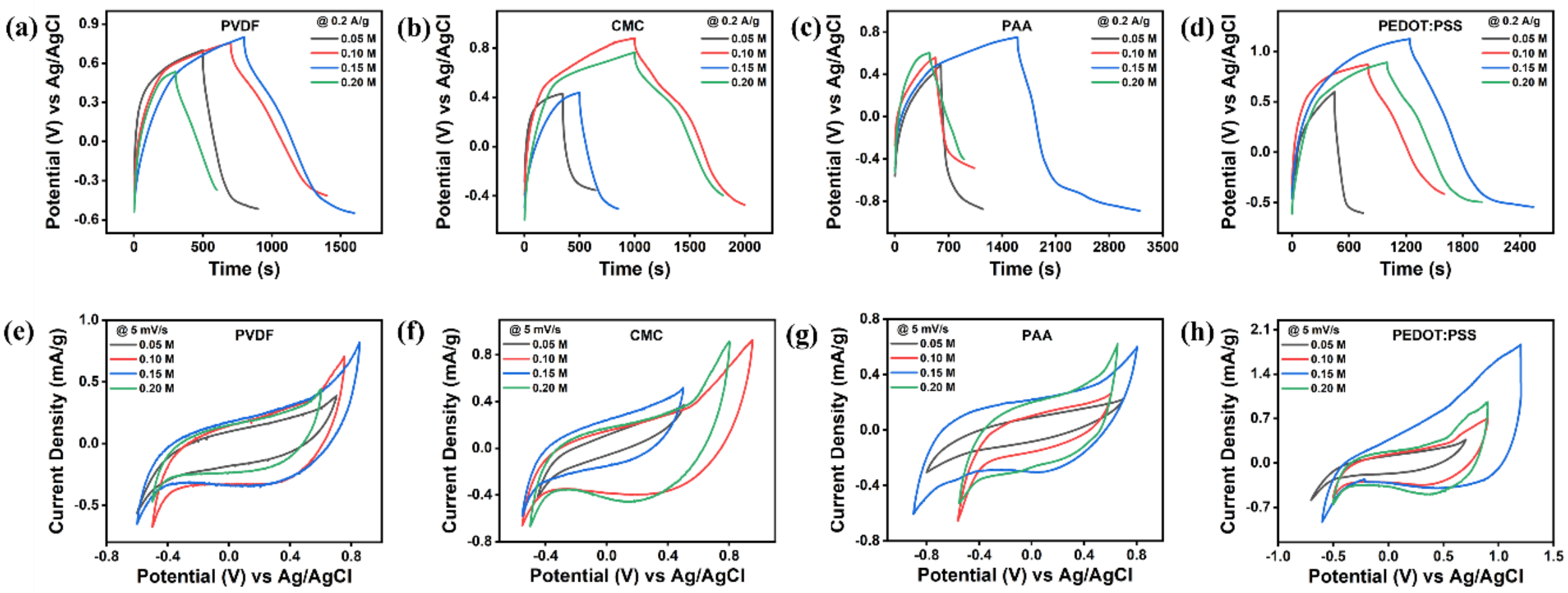

**Figure 3.** Electrochemical response of $CsPbBr_3$ electrodes at four LiTFSI concentrations in acetonitrile. (a–d) Galvanostatic charge–discharge profiles at 0.2 A g$^{-1}$ and (e–h) cyclic voltammograms at 5 mV s$^{-1}$ for electrodes prepared with (a, e) PVDF, (b, f) CMC, (c, g) PAA and (d, h) PEDOT:PSS. Potential windows differ between conditions and are listed in Table 1.

The voltammograms depart from the rectangular shape of an ideal double-layer capacitor. Each curve is tilted, indicating a resistive contribution, and each shows broad current features rather than a flat plateau, most clearly near −0.4 V in the CMC and PVDF panels. The galvanostatic profiles are correspondingly non-linear: charging is rapid at first and then curves toward the upper potential limit, and discharge shows an initial drop followed by an extended tail. Both features indicate that charge is stored by a combination of double-layer and faradaic processes rather than by electrostatic accumulation alone.[36] The relative contribution of the two mechanisms is not separated here.[37] Potential windows vary across the sixteen combinations, from 0.95 V for CMC at 0.05 M to 1.80 V for PEDOT:PSS at 0.15 M, and are listed in Table 1 alongside the derived quantities. Because energy density scales with the square of the potential window, values obtained at different windows are not directly comparable, and comparisons in this section are made on capacitance rather than energy density wherever the windows differ. Specific capacitance obtained from galvanostatic discharge at 0.2 A g$^{-1}$, areal capacitance obtained from voltammetry at 5 mV s$^{-1}$, and energy density are plotted against LiTFSI concentration in Figure 4a–c. The complete set of values is given in Table 1.

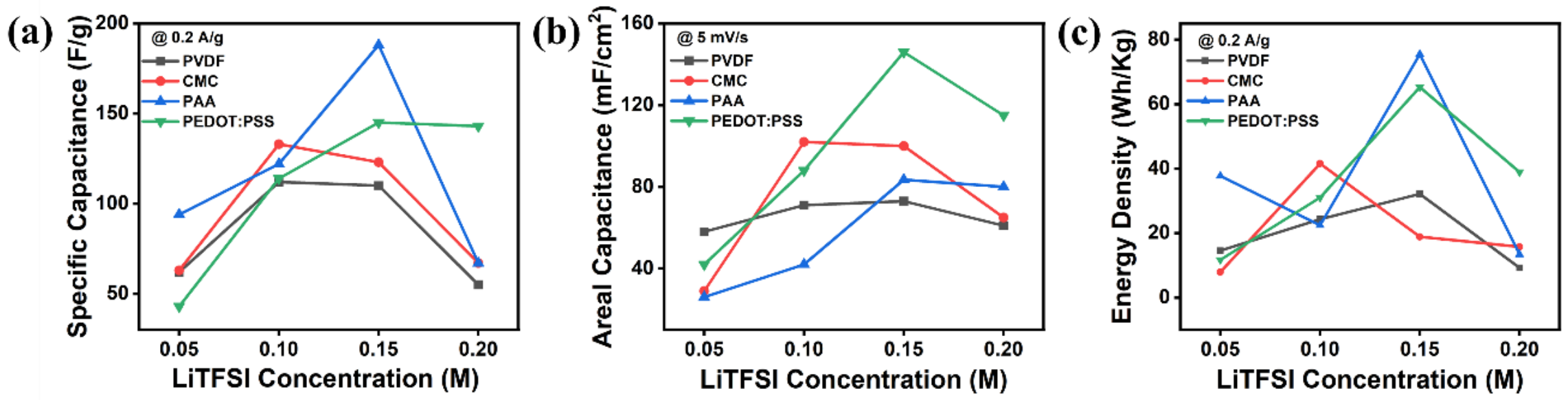

**Figure 4.** Dependence of electrochemical performance on LiTFSI concentration for all four binders. (a) Specific capacitance from galvanostatic discharge at 0.2 A $g^{-1}$, (b) areal capacitance from cyclic voltammetry at 5 mV $s^{-1}$, and (c) energy density calculated using the potential windows given in Table 1. All four binders reach maximum capacitance between 0.10 and 0.15 M and decline at 0.20 M.

All four binders show the same pattern with concentration. Capacitance rises from 0.05 M, reaches a maximum between 0.10 and 0.15 M, and falls at 0.20 M. For PVDF, specific capacitance increases from 62 F $g^{-1}$ at 0.05 M to 112 F $g^{-1}$ at 0.10 M, remains at 110 F $g^{-1}$ at 0.15 M and falls to 55 F $g^{-1}$ at 0.20 M. CMC follows the same course, increasing from 63 to 133 F $g^{-1}$ between 0.05 and 0.10 M, giving 123 F $g^{-1}$ at 0.15 M and 67 F $g^{-1}$ at 0.20 M. For both of these binders the values at 0.10 and 0.15 M differ by less than 10%, so the response is better described as a plateau across that range than as a maximum at a single concentration. PAA and PEDOT:PSS reach their highest values at 0.15 M. PAA gives 94, 122, 188 and 67 F $g^{-1}$ across the four concentrations, and the value of 188 F $g^{-1}$ at 0.15 M is the highest specific capacitance measured in this series. PEDOT:PSS gives 43, 114, 145 and 143 F $g^{-1}$, with the values at 0.15 and 0.20 M differing by less than 2%. Areal capacitance from voltammetry follows the same concentration dependence. PVDF gives 58, 71, 73 and 61 mF $cm^{-2}$, and CMC gives 29, 102, 100 and 65 mF $cm^{-2}$, both showing the plateau across 0.10–0.15 M seen in the gravimetric data. PAA gives 26, 42, 83.5 and 80 mF $cm^{-2}$ and PEDOT:PSS gives 42, 88, 146 and 115 mF $cm^{-2}$, both peaking at 0.15 M. The value of 146 mF $cm^{-2}$ for PEDOT:PSS at 0.15 M is the highest areal capacitance in the series. Equation (1) predicts a capacitance optimum at 0.10 M for all four binders at the 15 wt% loading used here. The measured optima are compared against this prediction in Table 3 of Section 4.1. The two measurements agree on the concentration dependence of each binder while ranking the binders differently against one another, since the gravimetric value normalises to active material mass and the areal value to geometric electrode area. The two are related through the mass loading, which was nominally 1 mg $cm^{-2}$ but was not measured individually for each electrode.[38] Energy densities in Figure 4c range from 7.9 Wh kg⁻¹ for CMC at 0.05 M to 75.5 Wh kg⁻¹ for PAA at 0.15 M. These values incorporate the potential window through equation (S5), and the highest values are obtained for the combinations with both high capacitance and wide windows: PAA at 0.15 M with 188 F g⁻¹ over 1.7 V, and PEDOT:PSS at 0.15 M with 145 F g⁻¹ over 1.8 V, giving 65.3 Wh kg⁻¹. Windows are reported alongside each value in Table 1 so that the two contributions can be separated. Two limits on the interpretation of these values should be stated. First, each condition was measured on a single electrode, so the differences quoted are not supported by replicate measurements, and differences below roughly 10% should not be treated as resolved. Second, the electrodes contain 15 wt% Super P, which contributes double-layer capacitance of its own, and a binder-and-carbon electrode without $CsPbBr_3$ was not measured. The reported values are therefore composite-electrode capacitances. The comparison between binders remains valid because carbon content, active material loading and electrode geometry are identical across all sixteen electrodes, but the fraction of the measured capacitance attributable to the perovskite is not established here.

**Table 1.** Potential window, discharge time at 0.2 A g⁻¹, specific capacitance, areal capacitance at 5 mV s⁻¹, energy density and charge transfer resistance for all sixteen binder–electrolyte combinations.

| **Binder** | **LiTFSI (M)** | **$\Delta V$ (V)** | **$C_s$ (F g⁻¹)** | **$C_A$ (mF cm⁻²)** | **$E$ (Wh kg⁻¹)** | **$R_{ct}$ (Ω)** |
|---|---|---|---|---|---|---|
| PVDF | 0.05 | 1.3 | 62 | 58 | 14.6 | 148 |
| | 0.10 | 1.25 | 112 | 71 | 24.3 | 43 |
| | 0.15 | 1.45 | 110 | 73 | 32.1 | 29 |
| | 0.20 | 1.1 | 55 | 61 | 9.2 | 19 |
| CMC | 0.05 | 0.95 | 63 | 29 | 7.9 | 75 |
| | 0.10 | 1.5 | 133 | 102 | 41.6 | 47 |
| | 0.15 | 1.3 | 123 | 100 | 28.9 | 30 |
| | 0.20 | 1.05 | 67 | 65 | 10.3 | 22.5 |
| PAA | 0.05 | 1.7 | 94 | 26 | 37.7 | 130 |
| | 0.10 | 1.15 | 122 | 42 | 22.4 | 46 |
| | 0.15 | 1.7 | 188 | 83.5 | 75.5 | 29 |
| | 0.20 | 1.2 | 67 | 80 | 13.4 | 29 |
| PEDOT:PSS | 0.05 | 1.4 | 43 | 42 | 11.7 | 180 |
| | 0.10 | 1.4 | 114 | 88 | 31 | 44 |
| | 0.15 | 1.8 | 145 | 146 | 65.3 | 30 |
| | 0.20 | 1.4 | 143 | 115 | 38.9 | 22 |

### 3.3 Impedance response

Nyquist plots for the four binders at all concentrations are shown in Figure 5a–d. Each spectrum consists of a depressed semicircle at high frequency, from which the charge transfer resistance is obtained, followed by a sloping low-frequency branch associated with ion diffusion. Fitted charge transfer resistances are plotted against concentration in Figure 5e and listed in Table 1; diffusion coefficients derived from the low-frequency response are shown in Figure 5f.

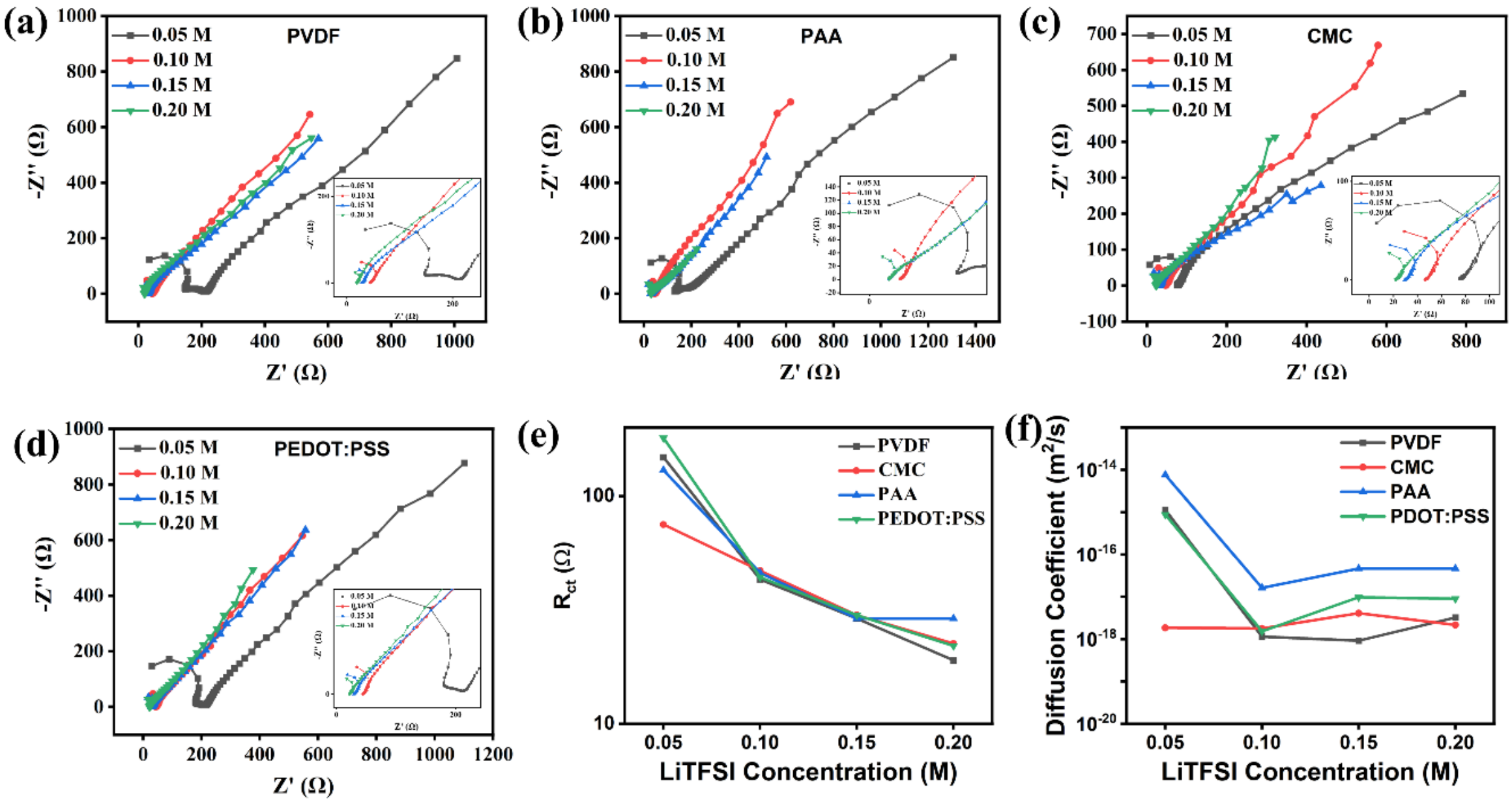

**Figure 5.** Electrochemical impedance response. (a–d) Nyquist plots at four LiTFSI concentrations for electrodes prepared with (a) PVDF, (b) PAA, (c) CMC and (d) PEDOT:PSS. (e) Fitted charge transfer resistance and (f) diffusion coefficient obtained from the Warburg region, both plotted against concentration on logarithmic axes. Charge transfer resistance decreases monotonically for every binder and converges to within 29–30 Ω at 0.15 M.

Charge transfer resistance decreases with increasing LiTFSI concentration for every binder. At 0.05 M the values differ widely between binders, from 75 Ω for CMC to 180 Ω for PEDOT:PSS, with PAA at 130 Ω and PVDF at 148 Ω. At 0.10 M all four fall to within 43–47 Ω, and at 0.15 M to within 29–30 Ω. Above 0.10 M the charge transfer resistance is therefore set by the electrolyte rather than by the binder, and the binder-dependent differences seen at 0.05 M do not persist. The concentration dependence of the resistance runs opposite to that of the capacitance. Charge transfer resistance is lowest at 0.20 M, at 19 Ω for PVDF, 22 Ω for PEDOT:PSS and 22.5 Ω for CMC, while capacitance at that concentration is the lowest measured for PVDF, CMC and PAA. Ion supply to the electrode surface therefore continues to improve as concentration increases beyond the capacitance maximum. The decline in stored charge above 0.15 M cannot be attributed to a transport limitation and must originate in a chemical or structural change at the electrode. PAA is the exception to the monotonic decrease. Its charge transfer resistance remains at 29 Ω between 0.15 and 0.20 M while the other three binders fall by 7–8 Ω over the same interval, and PAA also shows the largest capacitance loss at 0.20 M, from 188 to 67 F $g^{-1}$. The resistance therefore stops improving in the one system where the capacitance falls furthest, which is consistent with a change at the electrode interface rather than in the bulk electrolyte, though the present data do not identify its nature. Diffusion coefficients extracted from the Warburg region are shown in Figure 5f and span $1.85\times10^{-18}$ to $7.59\times10^{-15}$ $m^2$ $s^{-1}$ across the sixteen combinations. For PVDF, PAA and PEDOT:PSS the highest values occur at 0.05 M and drop by two to three orders of magnitude at 0.10 M, after which they change comparatively little; CMC shows no such drop and remains between $1.8\times10^{-18}$ and $4.1\times10^{-18}$ $m^2$ $s^{-1}$ throughout. The Warburg region is the least well-resolved part of the spectrum, and values extracted from it are more sensitive to the fitting range than the charge transfer resistance, so these numbers are reported as measured without being used to support the mechanistic argument.[31]

### 3.4 Structural evolution after electrochemical characterisation

Diffraction patterns recorded before and after the full electrochemical sequence are compared in Figure 6a–d, with the as-synthesised powder pattern reproduced in each panel for reference. Scanning electron micrographs of the PAA and PEDOT:PSS electrodes in both states are shown in Figure 6e–h, and the complete set for all four binders in Figure S9.

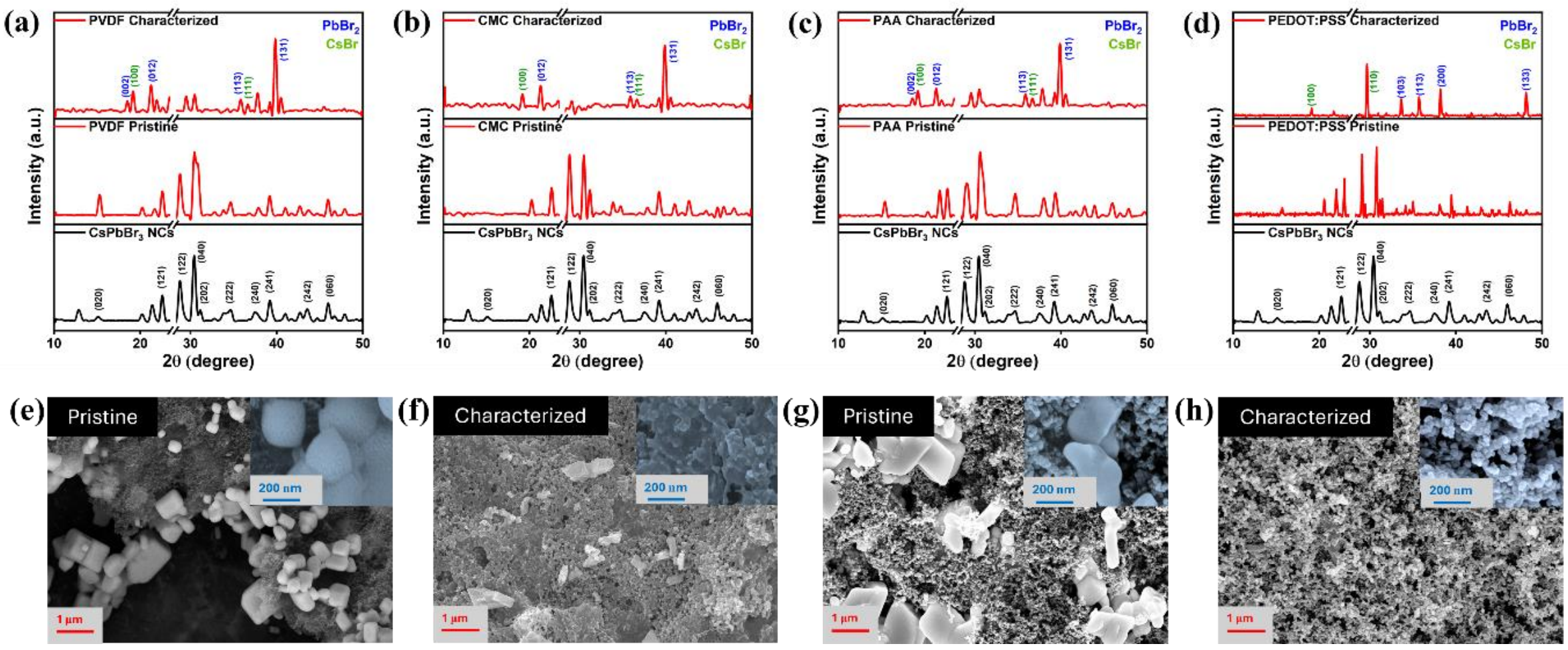

**Figure 6.** Structural and morphological change following electrochemical characterisation. (a–d) X-ray diffraction patterns of pristine and characterized electrodes prepared with (a) PVDF, (b) CMC, (c) PAA and (d) PEDOT:PSS, with the as-synthesised nanocrystal pattern reproduced from Figure 2a for comparison. Reflections of $PbBr_2$ (blue) and CsBr (green) appear after characterisation. (e–h) Field-emission scanning electron micrographs of (e, f) PAA and (g, h) PEDOT:PSS electrodes, (e, g) before and (f, h) after electrochemical characterisation.

In all four electrodes, the $CsPbBr_3$ reflections that dominate the pristine patterns are weak or absent after characterisation, and a new set of reflections appears in their place. These are indexed to orthorhombic $PbBr_2$ (JCPDS 00-005-0608) at 18.60° (002), 21.63° (012), 35.78° (113) and 39.66° (131), and to CsBr (JCPDS 00-005-0588) at 20.66° (100) and 36.36° (111). The products correspond to the decomposition of the perovskite into its binary constituents,

$$\mathrm{CsPbBr_3} \rightarrow \mathrm{CsBr} + \mathrm{PbBr_2} \qquad (4)$$

which is the route reported for $CsPbBr_3$ under electrical bias and under moisture exposure.[16] The $PbBr_2$ (131) reflection at 39.66° is the most intense feature in each of the three characterized patterns. Assignment of this reflection requires care, because $CsPbBr_3$ (241) falls at 39.24°, 0.42° away; the two are distinguished here by the simultaneous appearance of the remaining $PbBr_2$ reflections at 18.60°, 21.63° and 35.78°, none of which coincides with a $CsPbBr_3$ reflection, and by the concurrent loss of the $CsPbBr_3$ reflections at 28.66° and 30.45° that dominate the pristine patterns.

Identification of CsBr rests on two reflections. Its most intense line, (110) at 29.44°, overlaps the region containing the $CsPbBr_3$ (040) and (202) reflections and the $PbBr_2$ (013) reflection at 30.20°, and is not resolved in these patterns. Neither $Cs_4PbBr_6$ nor $CsPb_2Br_5$ was detected in the characterized electrodes; the $Cs_4PbBr_6$ reflection at 12.8° present in the as-synthesised powder does not appear in any electrode pattern,[39] and the $CsPb_2Br_5$ reflections at 11.7°, 18.8° and 23.9° were checked and not observed. The transformation is also visible in the electrode morphology. The pristine surfaces in Figure 6e and Figure 6g carry faceted crystals between roughly 200 nm and 1 μm across, sitting on a finer matrix of carbon and binder. After characterisation, Figure 6f and Figure 6h show these crystals largely gone, leaving a uniform fine particulate network. Since diffraction shows that crystalline $PbBr_2$ and CsBr are present after characterisation, the loss of the large crystals is not simply conversion into a different crystalline phase in place: material has been removed from the electrode surface. Dissolution of the decomposition products into the acetonitrile electrolyte is consistent with both observations, and with the loss of surface lead measured by photoelectron spectroscopy in Section 3.5. An alternative account, in which the large crystals remain but are covered by a finer reprecipitated layer that limits the escape depth of the photoelectron measurement, is not excluded by these data; distinguishing the two would require elemental analysis of the used electrolyte, which was not performed. The electrochemical sequence preceding these measurements lasted approximately 6 h and comprised galvanostatic cycling at six current densities, voltammetry at nine scan rates, and impedance spectroscopy at six bias potentials from 0.0 to 1.0 V. The impedance series holds the electrode at each bias while the frequency is swept to 0.01 Hz, so a substantial part of the total exposure is a potentiostatic hold at anodic potential rather than a cycling excursion. The four electrodes examined here were cycled over the windows listed in Table 1, which span 0.95 to 1.80 V, so the anodic limit reached differs between binders and the extent of conversion is compared across electrodes not held to a common upper potential. The decomposition observed here therefore follows from a combination of cycling and sustained anodic bias, and the present measurements do not separate the two contributions. These observations describe the electrode after a single characterisation sequence rather than after extended cycling, and no capacitance-retention measurement over repeated cycles is reported here. This structural change accounts for the impedance behaviour described in Section 3.3. Charge transfer resistance falls monotonically with LiTFSI concentration while capacitance falls above 0.15 M, so the loss of stored charge is not caused by restricted ion supply. Progressive conversion of $CsPbBr_3$ to $PbBr_2$ and CsBr, followed by dissolution of those products, removes active material from the electrode and provides a mechanism for the capacitance decline that operates independently of ion transport.

### 3.5 Surface chemistry

X-ray photoelectron spectra of the four pristine electrodes are shown in Figure 7a–c, and spectra of the PEDOT:PSS electrode before and after electrochemical characterisation in Figure 7d–f. Survey scans, Cs 3d spectra for all four binders and O 1s spectra for the three oxygen-containing binders are given in Figure S10. Binding energies were referenced to the adventitious C 1s line at 284.8 eV.

#### Pristine electrodes

The core levels of the perovskite appear at the positions expected for $CsPbBr_3$ in all four electrodes. Pb $4f_{7/2}$ and $4f_{5/2}$ are found at 137.95 and 142.82 eV, giving a spin–orbit separation of 4.87 eV; Br $3d_{5/2}$ and $3d_{3/2}$ at 68.62 and 69.40 eV; Cs $3d_{5/2}$ and $3d_{3/2}$ at 724.16 and 738.22 eV, separated by 14.06 eV; and Cs $4d_{5/2}$ and $4d_{3/2}$ at 75.53 and 77.88 eV.[16,40] The C 1s envelope in Figure 7c is fitted with components at 284.8 eV (C–C), 286.67 eV (C–N), 289.06 eV (O–C=O) and 291.15 eV ($CF_2$), the last assigned to the fluorinated carbon of PVDF and to $CF_2$ intensity in the PEDOT:PSS spectrum. The O–C=O component is present in the PAA and CMC spectra and absent from PVDF, confirming that the carboxyl-bearing binders are present within the sampling depth of the measurement.

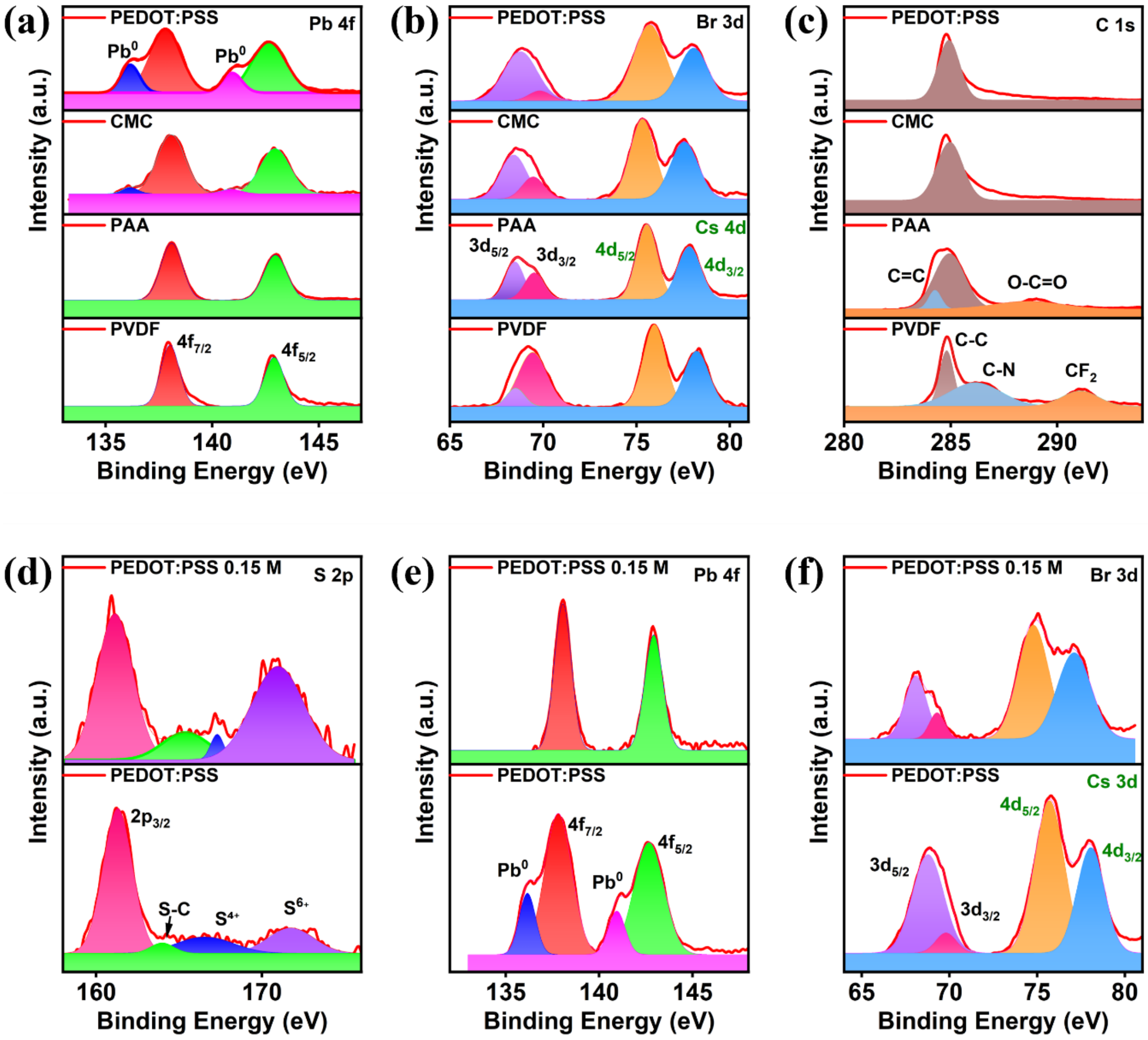

**Figure 7.** X-ray photoelectron spectra. (a) Pb 4f, (b) Br 3d with the Cs 4d doublet, and (c) C 1s regions for pristine electrodes prepared with all four binders. A component near 136.5 eV corresponding to metallic lead is present in the CMC and PEDOT:PSS spectra. (d) S 2p, (e) Pb 4f and (f) Br 3d regions for the PEDOT:PSS electrode before and after electrochemical characterisation in 0.15 M LiTFSI. Binding energies are referenced to the adventitious C 1s line at 284.8 eV.

Surface composition differs substantially between the four electrodes. Ratios of integrated peak areas are given in Table 2. These are area ratios rather than atomic ratios, since instrument transmission and sensitivity factors have not been applied; because the same correction applies to every sample, the ranking between binders is unaffected.

**Table 2.** Ratios of integrated XPS peak areas for the four pristine electrodes.

| Binder | Cs 3d / Pb 4f | Br 3d / Pb 4f | Br 3d / Cs 3d |
|---|---|---|---|
| PVDF | 1.82 | 1.58 | 0.87 |
| CMC | 3.81 | 1.93 | 0.51 |
| PAA | 3.58 | 1.17 | 0.33 |
| PEDOT:PSS | 5.04 | 1.37 | 0.27 |

Every surface is cesium-rich and bromide-poor relative to the bulk $CsPbBr_3$ stoichiometry, which is expected for nanocrystals terminated by cesium halide surfaces and passivated by oleylammonium bromide.[41] The extent differs by a factor of 2.8 between PVDF and PEDOT:PSS on the Cs/Pb ratio, and by a factor of 3.2 on the Br/Cs ratio. The ordering is the same for both quantities: PVDF is closest to stoichiometric, followed by CMC and PAA, with PEDOT:PSS furthest from it. The Pb 4f region in Figure 7a shows an additional component at low binding energy in two of the four electrodes. In the PEDOT:PSS spectrum this component is clearly resolved near 136.5 eV, the position of metallic lead,[40] and a weaker component at the same position is present in the CMC spectrum. No such component is detected for PVDF or PAA. Since no potential had been applied to these electrodes, the reduction of $Pb^{2+}$ occurred during slurry preparation or drying. X-ray induced reduction of $Pb^{2+}$ in halide perovskites during acquisition is documented[42] and must be considered, but all four samples contain the same perovskite and were measured on the same instrument, so beam damage would be expected to produce comparable signals in all four rather than in two. The two electrodes showing $Pb^0$ are also the two in which the binder does not dissolve in NMP. CMC is hygroscopic and PEDOT:PSS was introduced as an aqueous dispersion, so both introduce water into the slurry that PVDF and PAA do not; both also leave the nanocrystal surface less completely covered by polymer than a dissolved binder would. The present data do not distinguish between these two explanations, and doing so would require spectra of the dried binders alone, without perovskite, which were not recorded.

**After electrochemical characterisation**

Spectra of the PEDOT:PSS electrode after the full electrochemical sequence in 0.15 M LiTFSI are compared with the pristine spectra in Figure 7d–f. Equivalent spectra recorded after characterisation in 0.20 M LiTFSI were indistinguishable, indicating that the surface chemistry does not change measurably between these two concentrations.

The Pb 4f spectrum in Figure 7e shows two changes. The metallic lead component present in the pristine electrode is no longer detected, and the ratio of Pb 4f to Cs 3d integrated area falls from 0.199 to 0.122. Loss of the metallic component is expected: the $Pb^{2+}/Pb^{0}$ couple lies near −0.32 V against Ag/AgCl,[43] so metallic lead is unstable across almost the entire potential range used here and is oxidised on the first anodic excursion. The accompanying decrease in the Pb/Cs ratio indicates that lead is depleted from the surface region relative to cesium, consistent with the loss of material described in Section 3.4. The sulfur 2p region of the PEDOT:PSS electrode is fitted with four components in both states. The component at 164.0 eV in the pristine electrode corresponds to thiophene sulfur of the PEDOT backbone.[44] After characterisation this component shifts to 165.5 eV and broadens, and the intensity of the most oxidised component near 171 eV increases relative to the remainder of the spectrum. Both changes indicate oxidation of the thiophene sulfur to sulfoxide and sulfone species, the reported signature of PEDOT overoxidation, which proceeds irreversibly above approximately +0.8 V and breaks the conjugation of the polymer backbone.[45] The impedance sequence held the electrode at +0.8 and +1.0 V, so the conducting polymer degrades within the potential window used here. The position of the highest-binding-energy component, at 171.8 eV in the pristine electrode and 171.0 eV after characterisation, lies above the 168.3 eV expected for sulfonate sulfur; differential charging of the insulating PSS phase relative to the conductive PEDOT domains is the likely origin, so this component is assigned by oxidation state rather than by absolute position. The component at 161.1–161.2 eV lies in the range reported for sulfide sulfur[46,47] and is unchanged between the two states; its origin cannot be established from these data, since determining whether it arises from the polymer, from processing, or from interaction with lead would require a spectrum of the dried binder alone. Because F 1s and N 1s spectra were not recorded, a contribution from residual LiTFSI to the components between 167 and 171 eV cannot be excluded. Taken together with the diffraction results, the spectroscopic data show that PEDOT:PSS is the least chemically compatible of the four binders with $CsPbBr_3$ in this system. It produces the most cesium-rich and bromide-depleted surface, and the largest metallic lead fraction, and its own backbone is oxidised within the working potential window. The same electrode gives the highest areal capacitance in the series at 146 mF $cm^{-2}$.

# 4. Discussion

## 4.1 Binder dependence of the interfacial balance constant

The concentration optima measured here are compared in Table 3 against the prediction of equation (1) at the 15 wt% binder loading used throughout.

**Table 3.** Measured capacitance optimum, implied interfacial balance constant and normalised balance for each binder at 15 wt% loading. Predicted optimum from equation (1) with $\xi_{Int}$ = 25 is 0.10 M.

| **Binder** | **Optimum, specific** | **Optimum, areal** | $\xi_{Int}$ | $\boldsymbol{\lambda + \theta}$ |
|---|---|---|---|---|
| PVDF | 0.10 M | 0.15 M | 25–30 | 1.0–1.2 |
| CMC | 0.10 M | 0.10 M | 25 | 1 |
| PAA | 0.15 M | 0.15 M | 30 | 1.2 |
| PEDOT:PSS | 0.15 M | 0.15 M | 30 | 1.2 |

CMC matches the prediction on both measures. PVDF is consistent with it and gives an optimal value at 0.10–0.15 M: specific capacitance is 112 F $g^{-1}$ at 0.10 M and 110 F $g^{-1}$ at 0.15 M, and areal capacitance is 71 and 73 mF $cm^{-2}$ at the same two concentrations, so both quantities are flat across this interval and the position of the maximum cannot be assigned from a single electrode per condition. PAA and PEDOT:PSS both reach their maxima at 0.15 M on both measures, one concentration step above the prediction, giving $\xi_{Int}$ = 30 and $\lambda + \theta = 1.2$.

Three constraints limit what this comparison establishes. The concentration grid has 0.05 M resolution, so $\xi_{Int}$ values of 25 and 30 are separated by a single grid step and the difference is at the limit of what the measurement resolves. Only one binder loading was examined, so equation (6.1) can be tested at a single point for each binder but its slope cannot be determined independently. Each condition was measured on one electrode, so the 2–3% differences that make PVDF ambiguous are not resolved. The comparison establishes that two of the four binders deviate from the prediction in the same direction and by the same amount; it does not determine the functional form of the deviation.

Within those constraints, two results follow.

The relationship applies to $CsPbBr_3$. With PVDF at 15 wt%, the formulation used in the earlier study, the optimum falls at 0.10–0.15 M and gives 112 F $g^{-1}$, compared with 126 F $g^{-1}$ measured for $CsSnCl_3$ at the same loading and concentration.[26] $CsPbBr_3$ differs from the earlier systems in B-site cation, in halide, and in crystal system, being orthorhombic rather than cubic. The relationship therefore does not depend on the specific lattice, which extends the conclusion drawn previously from the comparison between $CsSnCl_3$ and $MASnCl_3$ where only the A-site cation was varied. The constant is not the same for all binders. If $\xi_{Int}$ depended only on the mass fraction of polymer, all four binders at 15 wt% would optimise at 0.10 M. PAA and PEDOT:PSS do not. The grouping does not follow the polymer backbone: CMC and PAA both carry carboxylate groups and fall on opposite sides. It does not follow electronic conductivity: PEDOT:PSS and PAA differ by orders of magnitude and group together. It does not follow the distribution state of the binder in the electrode: all four slurries were prepared in NMP, in which PVDF and PAA dissolve while CMC and PEDOT:PSS form dispersions, so each $\xi_{Int}$ group contains one dissolved and one dispersed binder. The two binders showing the shift are those carrying ionisable acid groups at high density along the chain. Equation (1) is expressed in weight percent, which does not distinguish between polymers delivering different numbers of functional groups per unit mass. Estimated ionisable group contents at 15 wt% loading are given in Table 4.

**Table 4.** Ionisable group content of the four binders at 15 wt% electrode loading.

| Binder | Repeat unit mass (g $mol^{-1}$) | Ionisable group | Group density (mmol $g^{-1}$ electrode) |
|---|---|---|---|
| PVDF | 64 | none | 0 |
| CMC | 210-234 (DS 0.6–0.9)[28] | $–COO^{-}$ | 0.43–0.58 |
| PAA | 72 | –COOH | 2.08 |
| PEDOT:PSS | 184 (PSS unit) | $–SO_3H$ | ~0.58 |

PVDF carries no ionisable groups. CMC carries carboxylates, but is supplied as the sodium salt, so its exchange sites are occupied by $Na^+$ before the slurry is made.[48,49] Both retain $\xi_{Int}$ = 25. PAA and PEDOT:PSS are supplied as free acids, with protons available for exchange, and both shift to 30. A mechanism consistent with this grouping is that free acid groups bind $Li^+$ within the polymer phase by ion exchange, so a higher bulk concentration is required to reach the same free $Li^+$ activity at the perovskite surface, displacing the optimum upward. Group density alone does not separate the two groups: at 15 wt% loading, CMC supplies 0.43–0.58 mmol $g^{-1}$ of carboxylate and PEDOT:PSS approximately 0.58 mmol $g^{-1}$ of sulfonate, values that overlap within the uncertainty of the degree of substitution and the PSS:PEDOT ratio. The ionisation state of the groups, rather than their number, correlates with the observed shift. This is offered as an interpretation rather than a demonstrated mechanism. With four binders measured at one loading, on a grid whose resolution equals the size of the shift, the data support a grouping and not a functional dependence. The interpretation is testable. Potentiometric titration of the exchangeable acid capacity of each binder, rather than its total group density, would establish whether the shift tracks the number of protons available for exchange. A direct test is available within a single binder: comparing carboxymethyl cellulose in its sodium and free acid forms at the same loading would vary the ionisation state while holding backbone, group density and molecular weight constant, and would show whether $\xi_{Int}$ shifts from 25 to 30 on protonation alone. Locating the optima on a finer concentration grid, across additional binders and additional loadings, would then determine whether $\xi_{Int}$ varies continuously or takes discrete values set by the ionisation state of the binder.

**4.2 Relationship between the interfacial optimum and lattice stability**

Diffraction after electrochemical characterisation showed no secondary phases in $CsSnCl_3$ or $MASnCl_3$ at any electrolyte concentration. In $CsPbBr_3$ the same measurement protocol produces $PbBr_2$ and CsBr in every electrode, as described in Section 3.4, and removes material from the electrode surface. The interfacial balance relationship holds in both cases. The concentration at which capacitance is maximised and the chemical stability of the perovskite under bias are therefore separate properties. The first is set at the polymer–electrolyte interface and follows equation (1); the second is a property of the lattice. A binder can position the optimum correctly without protecting the active material, and the present results show this directly: PAA gives the highest specific capacitance in the series at 188 F $g^{-1}$ and the electrode still converts to $PbBr_2$ and CsBr. Impedance supports this separation. Charge transfer resistance decreases monotonically with LiTFSI concentration for every binder and reaches its lowest values, 19–29 Ω, at 0.20 M, where capacitance is at its lowest for PVDF, CMC and PAA. Ion supply to the electrode improves as stored charge falls, so the decline above the optimum is not caused by restricted transport. Two processes therefore act in opposite directions as concentration increases: transport improves, and the accommodation capacity of the polymer-modified interface is exceeded once lithium supply passes the balance point, with additional lithium no longer contributing to reversible interfacial storage. The structural measurements of Section 3.4 identify a further loss channel operating in this system, in which conversion of $CsPbBr_3$ to $PbBr_2$ and CsBr followed by dissolution of those products removes active material altogether. The present data do not apportion the capacitance decline between these two contributions.

### 4.3 Binder-dependent chemical compatibility

Beyond its effect on the position of the optimum, the binder determines the chemical state of the perovskite before any potential is applied. Surface composition measured on the pristine electrodes varies by a factor of 2.8 in Cs/Pb and 3.2 in Br/Cs across the four binders, with PVDF closest to the bulk stoichiometry and PEDOT:PSS furthest from it. Metallic lead is detected in two of the four. The two binders showing surface reduction, CMC and PEDOT:PSS, are the two that do not dissolve in NMP. Both introduce water that PVDF and PAA do not — CMC through its hygroscopic backbone, PEDOT:PSS through the aqueous dispersion in which it is supplied — and both leave the nanocrystal surface less completely covered by polymer than a dissolved binder would.[50] These two explanations cannot be separated with the present data, and doing so would require spectra of the dried binders alone, which were not recorded. The consequence is a trade-off between performance and chemical compatibility. PEDOT:PSS gives the highest areal capacitance in the series at 146 mF $cm^{-2}$. The dispersion used here is specified as a high-conductivity grade, so electronic percolation through the binder is a plausible contribution to this value, although the contributions of binder conductivity and of the electrode's altered chemical state are not separated by these measurements. PEDOT:PSS also produces the most cesium-rich and bromide-depleted surface, the largest metallic lead fraction, while its own backbone oxidises within the potential window used here. Spectra after characterisation were recorded only for the PEDOT:PSS electrode, so comparable degradation of the other three binders cannot be excluded; PVDF in particular undergoes dehydrofluorination under reducing or strongly basic conditions.[51] PAA gives the highest specific capacitance at 188 F $g^{-1}$ with no detectable metallic lead in the pristine electrode. On the combined criteria of capacitance and chemical compatibility, PAA is the better choice among the four for $CsPbBr_3$ in this electrolyte.

## 5. Conclusion

Four polymer binders spanning fluorinated, carboxylic, cellulosic and sulfonic functional-group chemistry were compared on $CsPbBr_3$ nanocrystal electrodes at four LiTFSI concentrations in acetonitrile, with binder loading, active material loading, conductive carbon content, solvent and casting procedure held constant across the sixteen combinations. Capacitance passes through a maximum between 0.10 and 0.15 M for every binder and falls at 0.20 M. The highest specific capacitance in the series is 188 F $g^{-1}$ for PAA at 0.15 M, and the highest areal capacitance is 146 mF $cm^{-2}$ for PEDOT:PSS at the same concentration. The interfacial balance relationship established previously for lead-free tin halide perovskites applies to $CsPbBr_3$. With PVDF at 15 wt%, the optimum falls at 0.10–0.15 M and gives 112 F $g^{-1}$, against 126 F $g^{-1}$ measured for $CsSnCl_3$ under the same formulation and concentration.[26] $CsPbBr_3$ differs from the earlier systems in B-site cation, halide and crystal system, so the relationship does not depend on the specific lattice. The interfacial balance constant is binder-dependent. PVDF and CMC optimise at 0.10 M, giving $\xi_{Int}$ = 25 and reproducing the value obtained for the tin chloride systems, while PAA and PEDOT:PSS optimise one concentration step higher at 0.15 M, giving $\xi_{Int}$ = 30. The two binders showing the shift are those carrying ionisable acid groups at high density. Equation (1) therefore requires a binder-specific constant when expressed in weight percent, and the grouping suggests that expressing binder content as ionisable group density may recover a single value across chemistries. Testing this requires measurement of group density alongside optima determined on a finer concentration grid. Charge transfer resistance decreases monotonically with LiTFSI concentration for all four binders, from 75–180 Ω at 0.05 M to 19–29 Ω at 0.20 M, and reaches its lowest values where capacitance is poorest. Ion supply to the electrode surface continues to improve above the optimum, so the decline in stored charge is not caused by restricted transport. The binder determines the chemical state of the perovskite before any potential is applied. Surface composition on the pristine electrodes varies by a factor of 2.8 in Cs/Pb and 3.2 in Br/Cs across the four binders. Metallic lead is detected in the CMC and PEDOT:PSS electrodes and not in PVDF or PAA. The two binders showing surface reduction are the two that do not dissolve in NMP and that introduce water into the slurry. $CsPbBr_3$ converts to $PbBr_2$ and CsBr in all four electrodes during electrochemical characterisation, and the micron-scale crystals present in the pristine electrodes are largely absent afterwards. The surface Pb/Cs ratio of the PEDOT:PSS electrode falls from 0.199 to 0.122, and the thiophene sulfur of the PEDOT backbone is replaced by oxidised species, showing that the binder degrades within the potential window used here. The interfacial optimum and the chemical stability of the perovskite are therefore independent: PAA gives the highest specific capacitance in the series and its electrode converts to $PbBr_2$ and CsBr along with the rest. On the combined criteria of capacitance and chemical compatibility, PAA is the better choice among the four binders for $CsPbBr_3$ in LiTFSI/acetonitrile. More generally, these results identify binder chemistry, and not binder mass fraction alone, as the variable that positions the electrolyte optimum, while showing that no binder examined here prevents loss of active material during operation.

## References

1. Simon, P. & Gogotsi, Y. Materials for electrochemical capacitors. *Nat. Mater.* **7**, 845–854 (2008).

2. Conway, B. E. *Electrochemical Supercapacitors*. (Springer US, Boston, MA, 1999). doi:10.1007/978-1-4757-3058-6.

3. Augustyn, V., Simon, P. & Dunn, B. Pseudocapacitive oxide materials for high-rate electrochemical energy storage. *Energy Environ. Sci.* **7**, 1597 (2014).

4. Eames, C. *et al.* Ionic transport in hybrid lead iodide perovskite solar cells. *Nat. Commun.* **6**, 7497 (2015).

5. Azpiroz, J. M., Mosconi, E., Bisquert, J. & De Angelis, F. Defect migration in methylammonium lead iodide and its role in perovskite solar cell operation. *Energy Environ. Sci.* **8**, 2118–2127 (2015).

6. Yuan, Y. & Huang, J. Ion Migration in Organometal Trihalide Perovskite and Its Impact on Photovoltaic Efficiency and Stability. *Acc. Chem. Res.* **49**, 286–293 (2016).

7. Manser, J. S., Christians, J. A. & Kamat, P. V. Intriguing Optoelectronic Properties of Metal Halide Perovskites. *Chem. Rev.* **116**, 12956–13008 (2016).

8. Protesescu, L. *et al.* Nanocrystals of Cesium Lead Halide Perovskites ($CsPbX_3$, X = Cl, Br, and I): Novel Optoelectronic Materials Showing Bright Emission with Wide Color Gamut. *Nano Lett.* **15**, 3692–3696 (2015).

9. Rout, C. S., Shinde, P., Belgami, M. A., Cho, J. S. & Jeong, S. M. Halide Perovskites for Supercapacitors and Photosupercapacitors: Recent Developments and Future Perspectives. *Small* **21**, (2025).

10. Kostopoulou, A., Brintakis, K., Nasikas, N. K. & Stratakis, E. Perovskite nanocrystals for energy conversion and storage. *Nanophotonics* **8**, 1607–1640 (2018).

11. Güz, S., Buldu-Akturk, M., Göçmez, H. & Erdem, E. All-in-One Electric Double Layer Supercapacitors Based on $CH_3NH_3PbI_3$ Perovskite Electrodes. *ACS Omega* **7**, 47306–47316 (2022).

12. Kumar, R. & Bag, M. Quantifying Capacitive and Diffusion-Controlled Charge Storage from 3D Bulk to 2D Layered Halide Perovskite-Based Porous Electrodes for Efficient Supercapacitor Applications. *The Journal of Physical Chemistry C* **125**, 16946–16954 (2021).

13. Kumar, T., Kumar, M., Kumar, A., Kumar, R. & Bag, M. A Review of Current Progress in Perovskite-Based Energy Storage to Photorechargeable Systems. *Energy & Fuels* **39**, 9185–9231 (2025).

14. Samatov, M. R. *et al.* Ion Migration at Metal Halide Perovskite Grain Boundaries Elucidated with a Machine Learning Force Field. *J. Phys. Chem. Lett.* **15**, 12362–12369 (2024).

15. Thiesbrummel, J. *et al.* Ion migration in perovskite solar cells. *Nat. Rev. Chem.* **10**, 179–195 (2026).

16. Kumar, A., Suhail, A., Sagar Shukla, P. & Bag, M. Structural Stability of Mixed-Halide Perovskite Nanocrystals in Energy Storage: The Role of Iodine Expulsion. *ChemNanoMat* **10**, (2024).

17. Prasanna, R. *et al.* Band Gap Tuning via Lattice Contraction and Octahedral Tilting in Perovskite Materials for Photovoltaics. *J. Am. Chem. Soc.* **139**, 11117–11124 (2017).

18. Dipta, S. S., Rahim, Md. A. & Uddin, A. Encapsulating perovskite solar cells for long-term stability and prevention of lead toxicity. *Appl. Phys. Rev.* **11**, (2024).

19. Srivastava, M., M. R., A. K. & Zaghib, K. Binders for Li-Ion Battery Technologies and Beyond: A Comprehensive Review. *Batteries* **10**, 268 (2024).

20. Bresser, D., Buchholz, D., Moretti, A., Varzi, A. & Passerini, S. Alternative binders for sustainable electrochemical energy storage – the transition to aqueous electrode processing and bio-derived polymers. *Energy Environ. Sci.* **11**, 3096–3127 (2018).

21. Qin, T., Yang, H., Li, Q., Yu, X. & Li, H. Design of functional binders for high-specific-energy lithium-ion batteries: from molecular structure to electrode properties. *Industrial Chemistry & Materials* **2**, 191–225 (2024).

22. Nguyen, V. A. & Kuss, C. Review—Conducting Polymer-Based Binders for Lithium-Ion Batteries and Beyond. *J. Electrochem. Soc.* **167**, 065501 (2020).

23. Nugraha, I. M. *et al.* An Alternative Polymer Material to PVDF Binder and Carbon Additive in Li-Ion Battery Positive Electrode. *Advanced Science* **11**, (2024).

24. Travis, W., Glover, E. N. K., Bronstein, H., Scanlon, D. O. & Palgrave, R. G. On the application of the tolerance factor to inorganic and hybrid halide perovskites: a revised system. *Chem. Sci.* **7**, 4548–4556 (2016).

25. De Roo, J. *et al.* Highly Dynamic Ligand Binding and Light Absorption Coefficient of Cesium Lead Bromide Perovskite Nanocrystals. *ACS Nano* **10**, 2071–2081 (2016).

26. Kumar, A. *et al.* Universality of PVDF-Li$^+$ Ion Interface Chemistry in Lead-free Perovskite Energy Storage Devices. https://arxiv.org/abs/2608.08295v1 (2026).

27. Konstantakou, M. & Stergiopoulos, T. A critical review on tin halide perovskite solar cells. *J. Mater. Chem. A Mater.* **5**, 11518–11549 (2017).

28. Kauling, J., Vankani, C., Winter, M. & Börner, M. Influence of the Degree of Substitution of Carboxymethyl Cellulose Binders on the Properties and Performance of Aqueously Processed $LiNi_{0.6}Mn_{0.2}Co_{0.2}O_2$-Based Positive Electrodes—A Comparative Study. *Advanced Energy and Sustainability Research* **7**, (2026).

29. Li, X. *et al.* CsPbX3 Quantum Dots for Lighting and Displays: Room-Temperature Synthesis, Photoluminescence Superiorities, Underlying Origins and White Light-Emitting Diodes. *Adv. Funct. Mater.* **26**, 2435–2445 (2016).

30. Wu, F. *et al.* High-Mass-Loading Electrodes for Advanced Secondary Batteries and Supercapacitors. *Electrochemical Energy Reviews* **4**, 382–446 (2021).

31. Nguyen, T. Q. & Breitkopf, C. Determination of Diffusion Coefficients Using Impedance Spectroscopy Data. *J. Electrochem. Soc.* **165**, E826–E831 (2018).

32. Stoumpos, C. C. *et al.* Crystal Growth of the Perovskite Semiconductor $CsPbBr_3$: A New Material for High-Energy Radiation Detection. *Cryst. Growth Des.* **13**, 2722–2727 (2013).

33. Akkerman, Q. A. *et al.* Nearly Monodisperse Insulator Cs4PbX6 (X = Cl, Br, I) Nanocrystals, Their Mixed Halide Compositions, and Their Transformation into $CsPbX_3$ Nanocrystals. *Nano Lett.* **17**, 1924–1930 (2017).

34. Protesescu, L. *et al.* Nanocrystals of Cesium Lead Halide Perovskites (CsPbX3 , X = Cl, Br, and I): Novel Optoelectronic Materials Showing Bright Emission with Wide Color Gamut. *Nano Lett.* **15**, 3692–3696 (2015).

35. Liu, M. *et al.* Unveiling Solvent-Related Effect on Phase Transformations in CsBr–$PbBr_2$ System: Coordination and Ratio of Precursors. *Chemistry of Materials* **30**, 5846–5852 (2018).

36. Brousse, T., Bélanger, D. & Long, J. W. To Be or Not To Be Pseudocapacitive? *J. Electrochem. Soc.* **162**, A5185–A5189 (2015).

37. Harish, S. & Sathyakam, P. U. Dunn's Method for Distinguishing Charge Storage Mechanisms in Supercapacitors: A Status Quo Review. *J. Electron. Mater.* **54**, 10858–10872 (2025).

38. Zahorodna, V. *et al.* Increasing Specific Capacitance by Optimization of the Thickness of Carbon Electrodes. *Batter. Supercaps* **8**, (2025).

39. Palazon, F. *et al.* Postsynthesis Transformation of Insulating $Cs_4PbBr_6$ Nanocrystals into Bright Perovskite $CsPbBr_3$ through Physical and Chemical Extraction of CsBr. *ACS Energy Lett.* **2**, 2445–2448 (2017).

40. Qaid, S. M. H., Ghaithan, H. M., Al-Asbahi, B. A. & Aldwayyan, A. S. Achieving Optical Gain of the CsPbBr3 Perovskite Quantum Dots and Influence of the Variable Stripe Length Method. *ACS Omega* **6**, 5297–5309 (2021).

41. Ravi, V. K. *et al.* Origin of the Substitution Mechanism for the Binding of Organic Ligands on the Surface of $CsPbBr_3$ Perovskite Nanocubes. *J. Phys. Chem. Lett.* **8**, 4988–4994 (2017).

42. Béchu, S., Ralaiarisoa, M., Etcheberry, A. & Schulz, P. Photoemission Spectroscopy Characterization of Halide Perovskites. *Adv. Energy Mater.* **10**, (2020).

43. Lebègue, E. Allen J. Bard, Larry. R. Faulkner, Henry S. White: Electrochemical Methods: Fundamentals and Applications, 3rd edition, Wiley. *Transition Metal Chemistry* **48**, 433–436 (2023).

44. Greczynski, G., Kugler, T. & Salaneck, W. R. Characterization of the PEDOT-PSS system by means of X-ray and ultraviolet photoelectron spectroscopy. *Thin Solid Films* **354**, 129–135 (1999).

45. TEHRANI, P. *et al.* The effect of pH on the electrochemical over-oxidation in PEDOT:PSS films. *Solid State Ion.* **177**, 3521–3527 (2007).

46. Gerson, A. R. & Bredow, T. Interpretation of sulphur 2p XPS spectra in sulfide minerals by means ofab initio calculations. *Surface and Interface Analysis* **29**, 145–150 (2000).

47. Krylova, V. & Andrulevičius, M. Optical, XPS and XRD Studies of Semiconducting Copper Sulfide Layers on a Polyamide Film. *International Journal of Photoenergy* **2009**, (2009).

48. Wei, L., Chen, C., Hou, Z. & Wei, H. Poly (acrylic acid sodium) grafted carboxymethyl cellulose as a high performance polymer binder for silicon anode in lithium ion batteries. *Sci. Rep.* **6**, 19583 (2016).

49. Mathew, A. *et al.* Limitations of Polyacrylic Acid Binders When Employed in Thick LNMO Li-ion Battery Electrodes. *J. Electrochem. Soc.* **171**, 020531 (2024).

50. Fanizza, E. *et al.* CsPbBr3 Nanocrystals-Based Polymer Nanocomposite Films: Effect of Polymer on Spectroscopic Properties and Moisture Tolerance. *Energies (Basel).* **13**, 6730 (2020).

51. Castillo, J. *et al.* Dehydrofluorination Process of Poly(vinylidene difluoride) PVdF-Based Gel Polymer Electrolytes and Its Effect on Lithium-Sulfur Batteries. *Gels* **9**, 336 (2023).

## *Supporting Information*

# Binder chemistry sets the interfacial balance constant in $CsPbBr_3$ nanocrystal supercapacitor electrodes

*Arun Kumar*[†], *Monojit Bag*[†, ‡, *].

†Advanced Research in Electrochemical Impedance Spectroscopy Laboratory, Indian Institute of Technology Roorkee, Roorkee 247667, India

‡Centre for Nanotechnology, Indian Institute of Technology Roorkee, Roorkee 247667, India

## Calculations

**Areal ($C_A$)** and **specific ($C_S$) capacitance** calculated from CV measurements using these equations.

$$C_A = \frac{\int(IdV)}{V\times s\times A} \tag{S1}$$

$$C_s = \frac{\int(IdV)}{V\times s\times m} \tag{S2}$$

**Areal ($C_A$)** and **specific ($C_S$) capacitance** calculated from GCD measurements using these equations.

$$C_A = \frac{I\int(Vdt)}{m\times V^2} \tag{S3}$$

$$C_s = \frac{I\int(Vdt)}{A\times V^2} \tag{S4}$$

Energy density and power density calculated as

$$E = \frac{1}{7.2}\times C_s(V)^2 \tag{S5}$$

$$P = \frac{E}{t} \tag{S6}$$

where ∫ (IdV) is the active area under CV spectra, V is potential window (V), s is the scan rate (mV $s^{-1}$), A is the active electrode area, m is mass of the active material (g) and t is the galvanostatic discharging time.

**Bandara-Mellander (B-M) Formalism**

The dielectric loss tangent can be calculated directly from electrochemical impedance data using this following relation:

$$\tan(\phi) = \frac{Z'}{Z''} \qquad \text{(S7)}$$

The Bandara and Mellander (B-M) model, which considers the effect of space charge polarization in its estimation of ionic diffusion coefficient, is used for that purpose. Using the B-M model, the relaxation time for the space charge and all other necessary parameters are calculated using either the loss tangent plot or the boundary condition of the dielectric function. This has been widely used by various research groups in recent investigations of the metal halide perovskite (MHP)-based optoelectronic device. Using this formulation, both real ($\epsilon'$) and imaginary ($\epsilon''$) parts of the complex dielectric constant, as well as the dielectric loss: tangent, can be represented as:

$$\epsilon' = \epsilon'_\infty \left(1 + \frac{\delta}{1 + (\omega\tau_1\delta)^2}\right) \quad \text{(S8)}$$

$$\epsilon'' = \epsilon'_\infty \left(1 + \frac{\omega\tau_1\delta^2}{1 + (\omega\tau_1\delta)^2}\right) \quad \text{(S9)}$$

$$\tan(\phi) = \frac{\epsilon''}{\epsilon'} = \frac{\omega\tau_1\delta}{1 + \omega^2\tau_1^2\delta} \quad \text{(S10)}$$

Here, $\tau_1$ represents the recombination time, which is related to the macroscopic relaxation time ($\tau_2$) through the following equation:

$$\tau_2 = \tau_1\sqrt{\delta} \quad \text{(S11)}$$

The point of maximum of the loss tangent ($\tan(\phi)$) function is directly proportional to $\tau_2$, whereas the parameter $\delta$ is derived from the loss tangent maximum. Hence, the loss tangent spectrum obtained experimentally can be described through the $\tan(\phi)$ expression as described above.

Finally, by means of B-M method, the macroscopic ion diffusion coefficient ($D_i$) is expressed as:

$$D_i = \frac{L^2}{\tau_2 \delta^2} \quad \text{(S12)}$$

where $L$ denotes the thickness of the electrode.

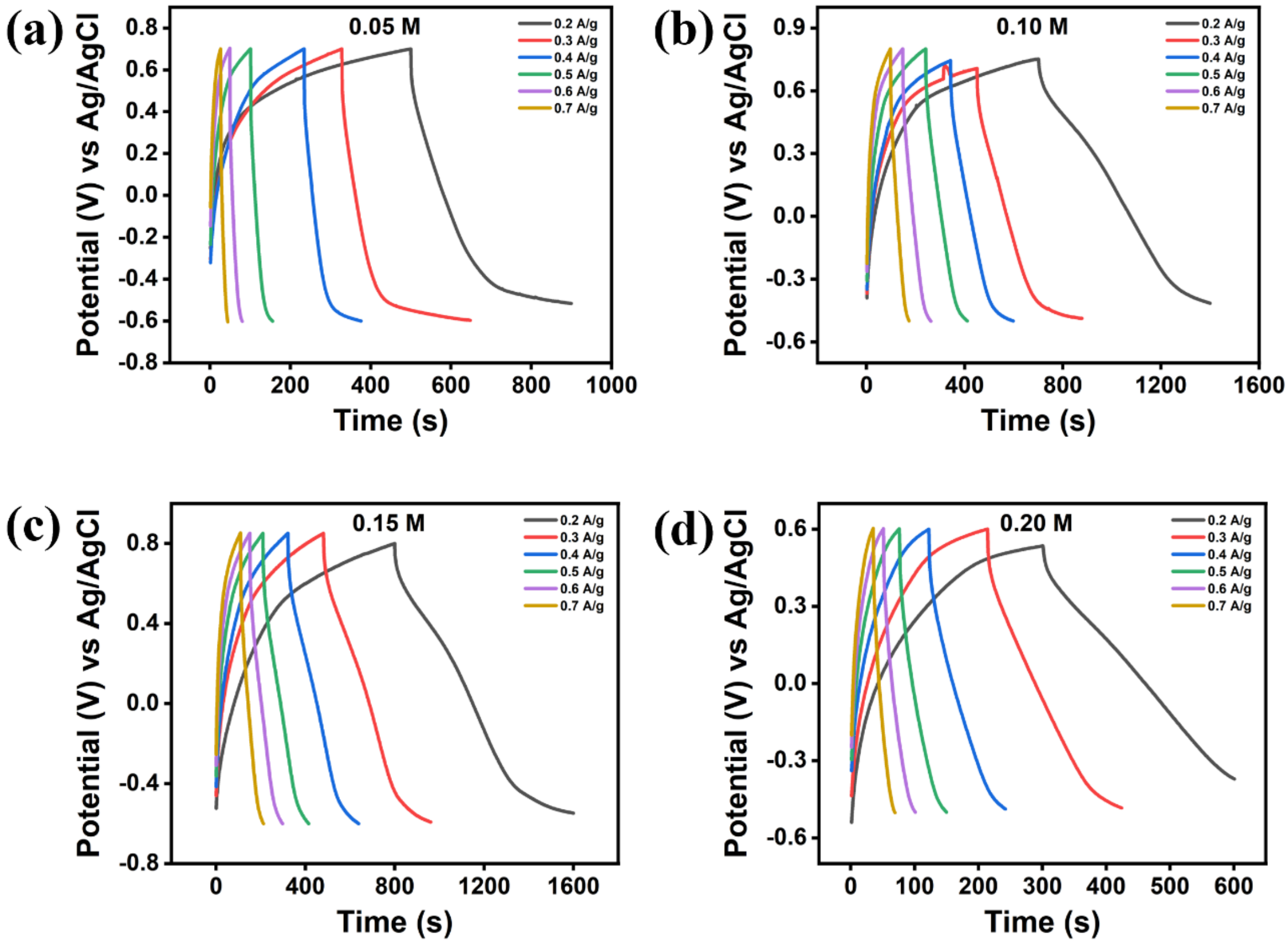

**Figure S1.** Galvanostatic charge–discharge profiles at 0.2–0.7 A $g^{-1}$ for PVDF binder at all four LiTFSI concentrations.

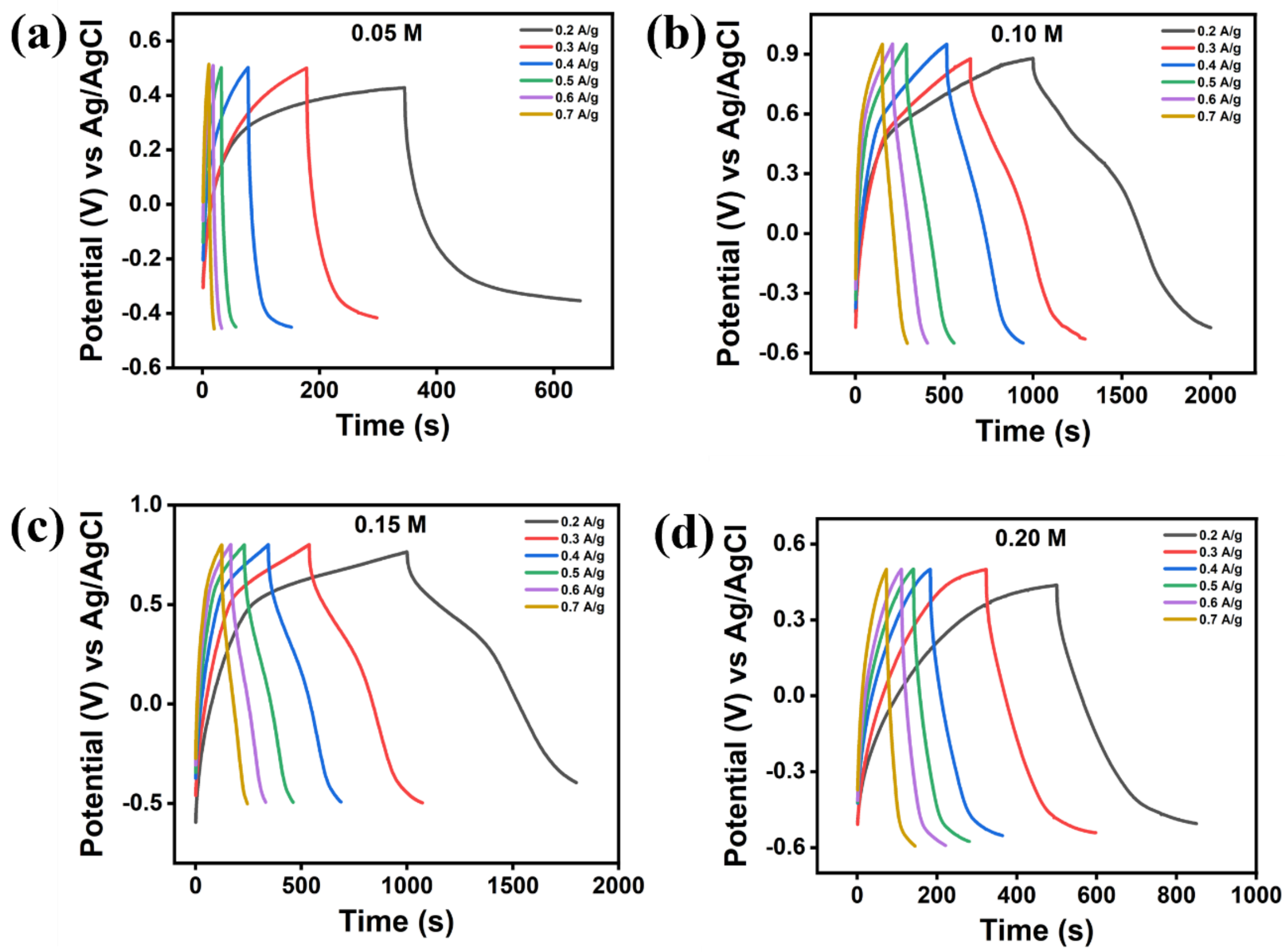

**Figure S2.** Galvanostatic charge–discharge profiles at 0.2–0.7 A $g^{-1}$ for CMC binder at all four LiTFSI concentrations.

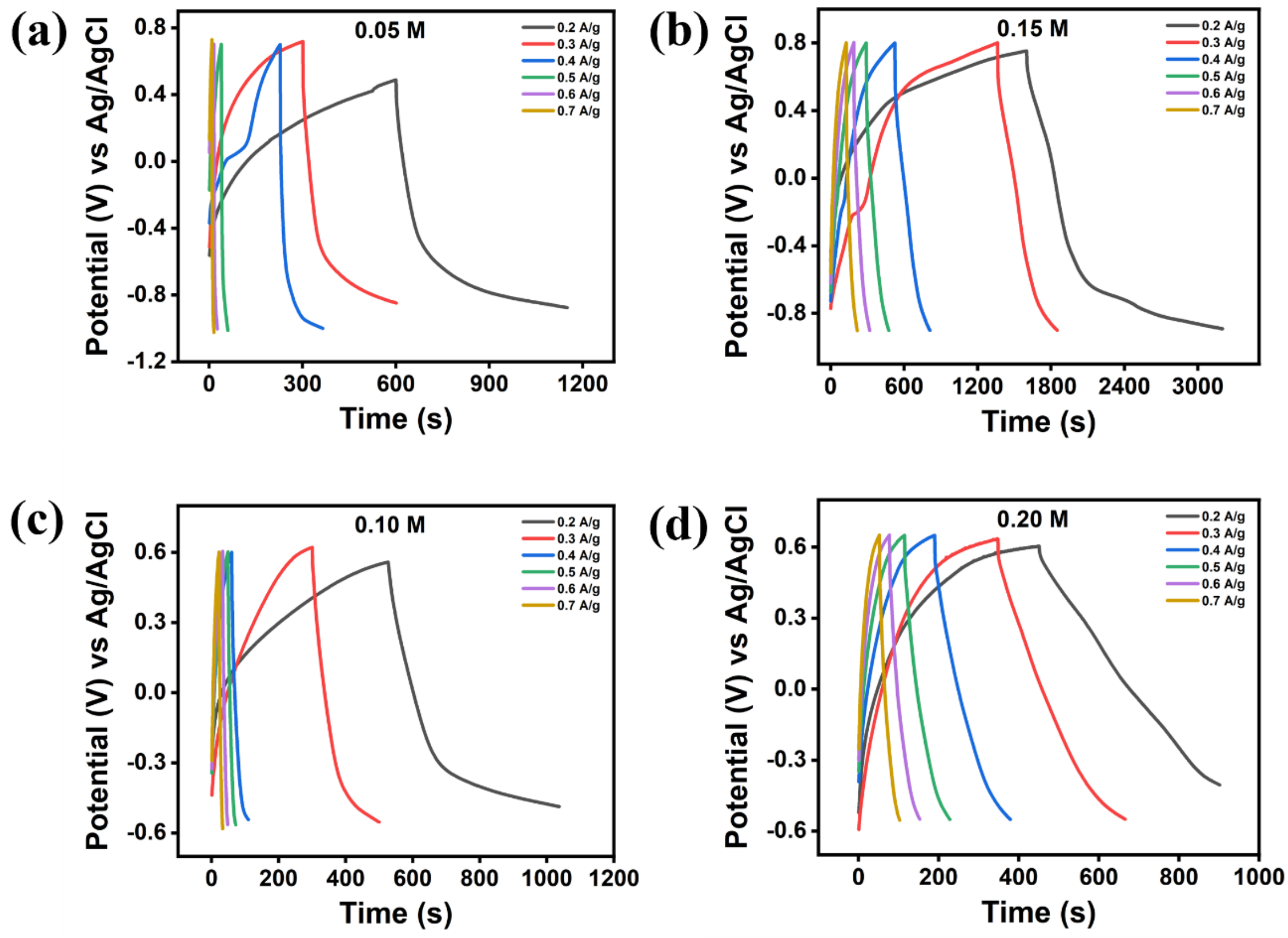

**Figure S3.** Galvanostatic charge–discharge profiles at 0.2–0.7 A $g^{-1}$ for PAA binder at all four LiTFSI concentrations.

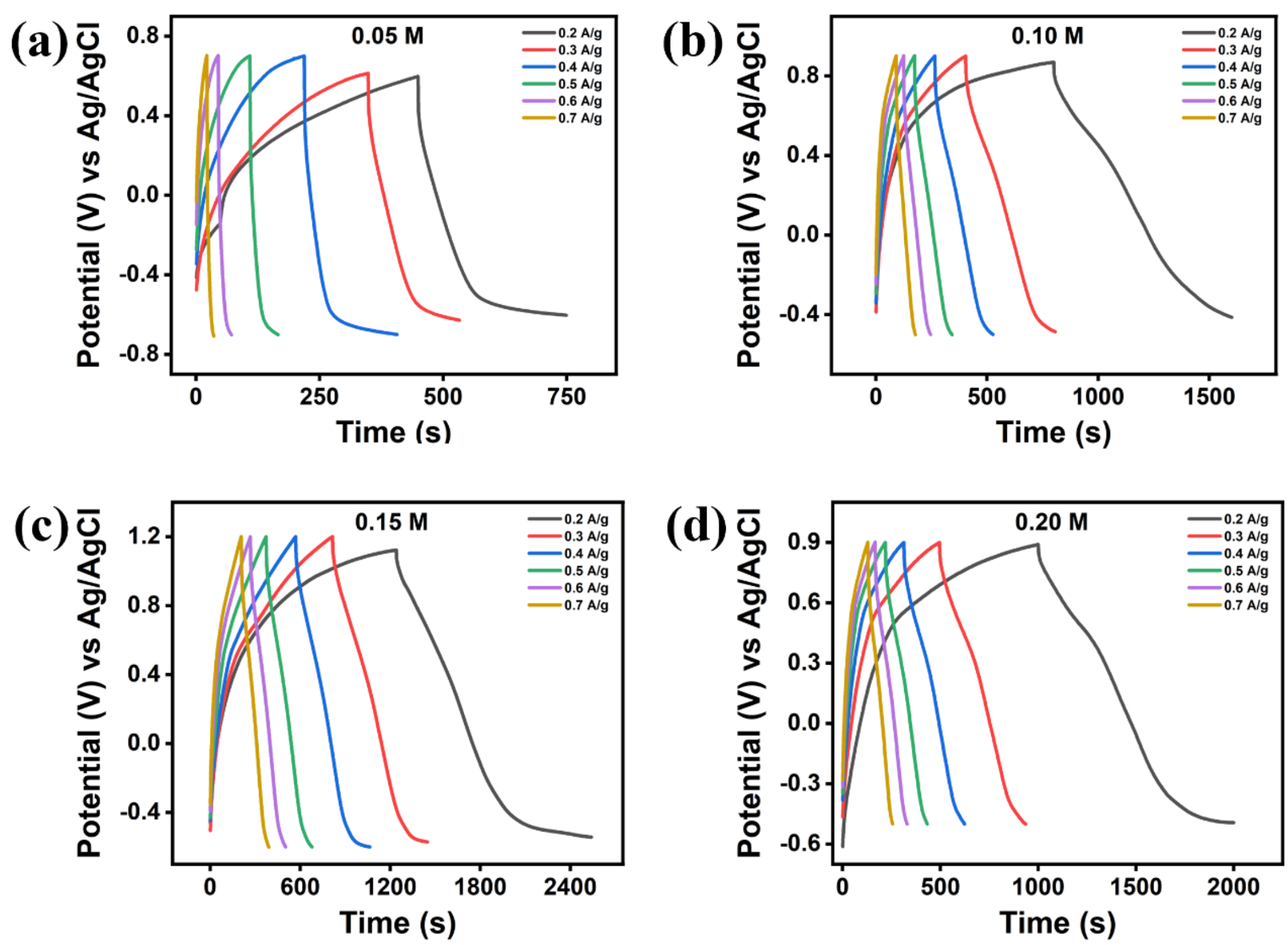

**Figure S4.** Galvanostatic charge–discharge profiles at 0.2–0.7 A g⁻¹ for PEDOT:PSS binder at all four LiTFSI concentrations.

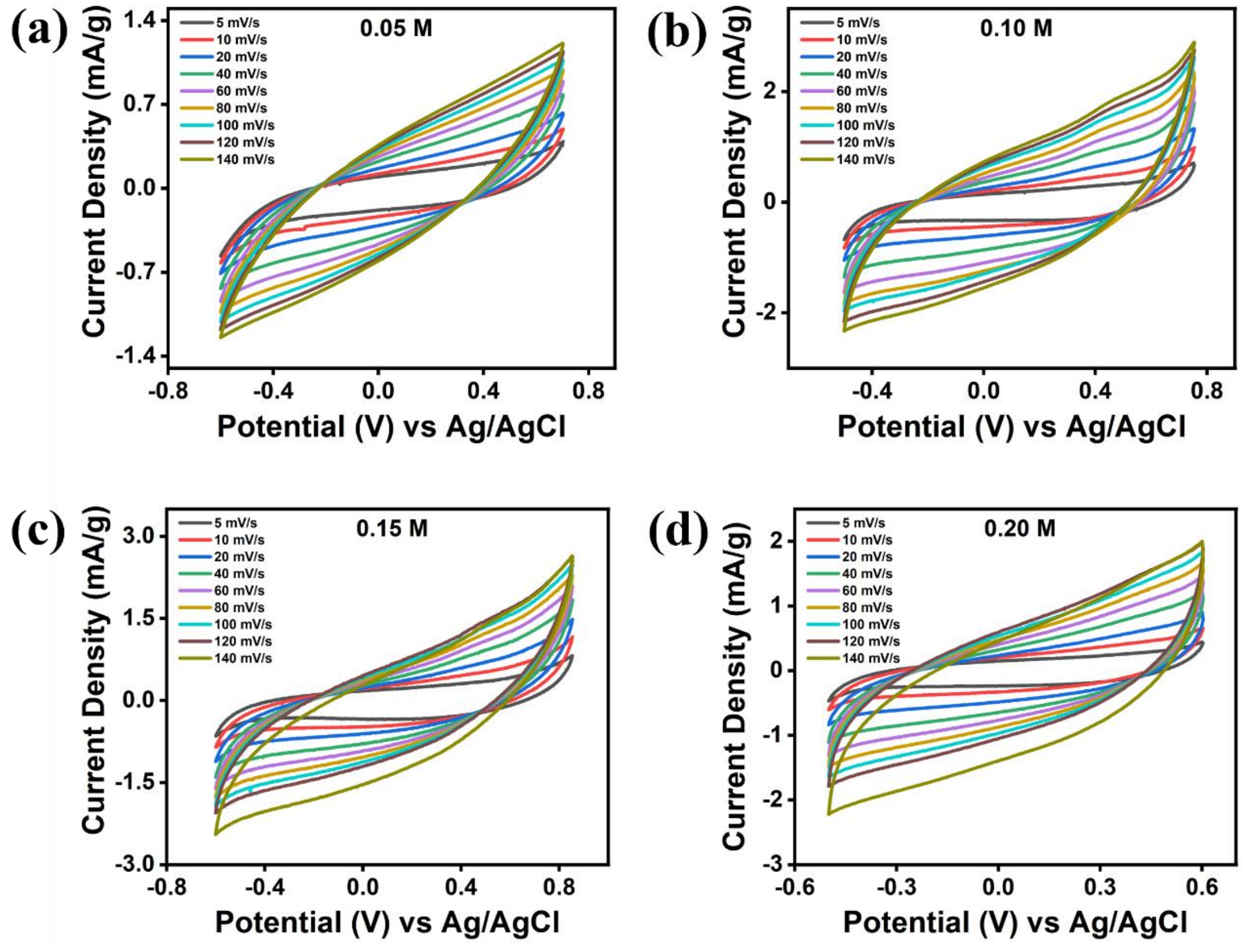

**Figure S5.** Cyclic voltammograms at 5–140 mV s$^{-1}$ for PVDF binder at all four LiTFSI concentrations.

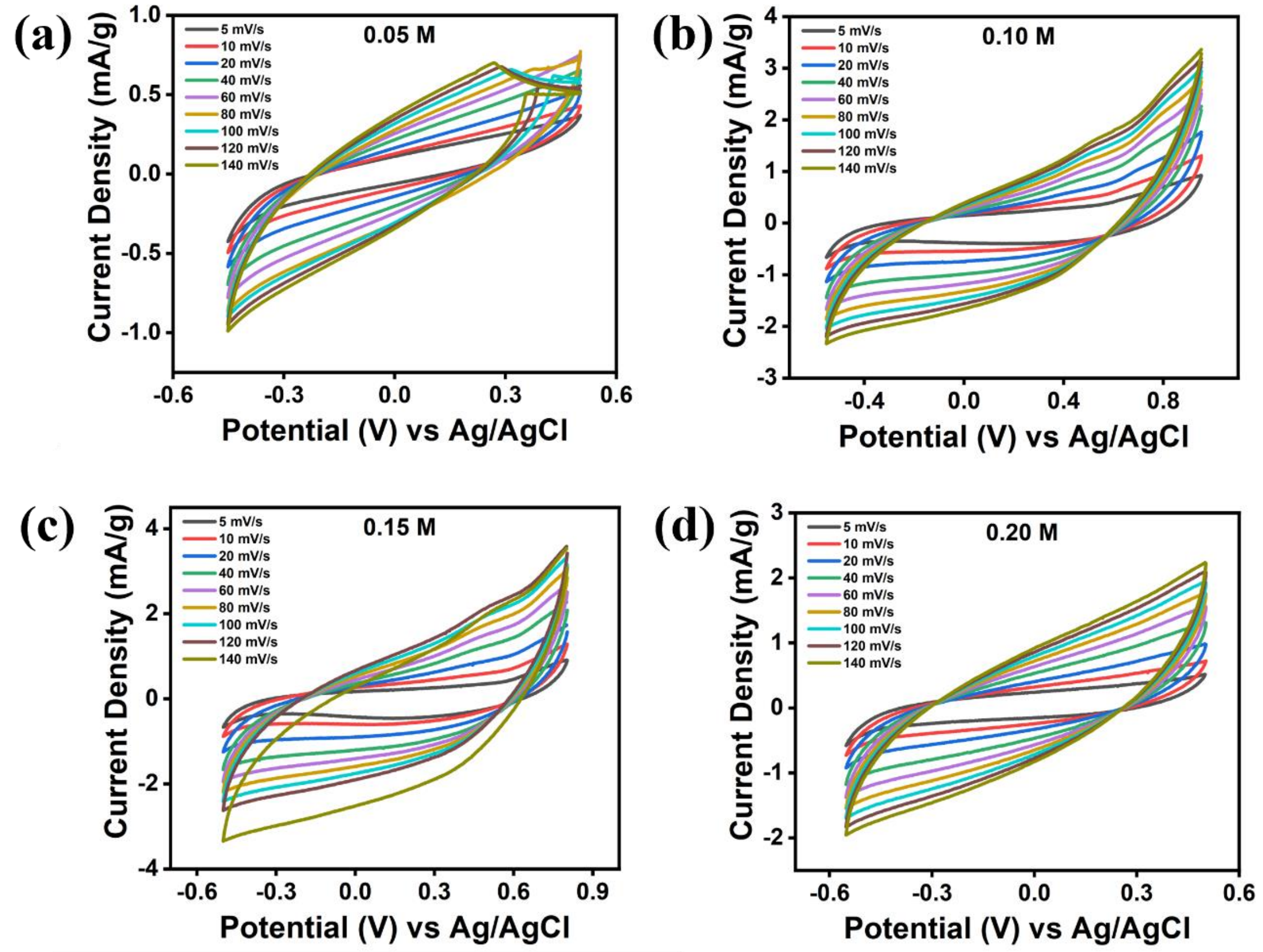

**Figure S6.** Cyclic voltammograms at 5–140 mV $s^{-1}$ for CMC binder at all four LiTFSI concentrations.

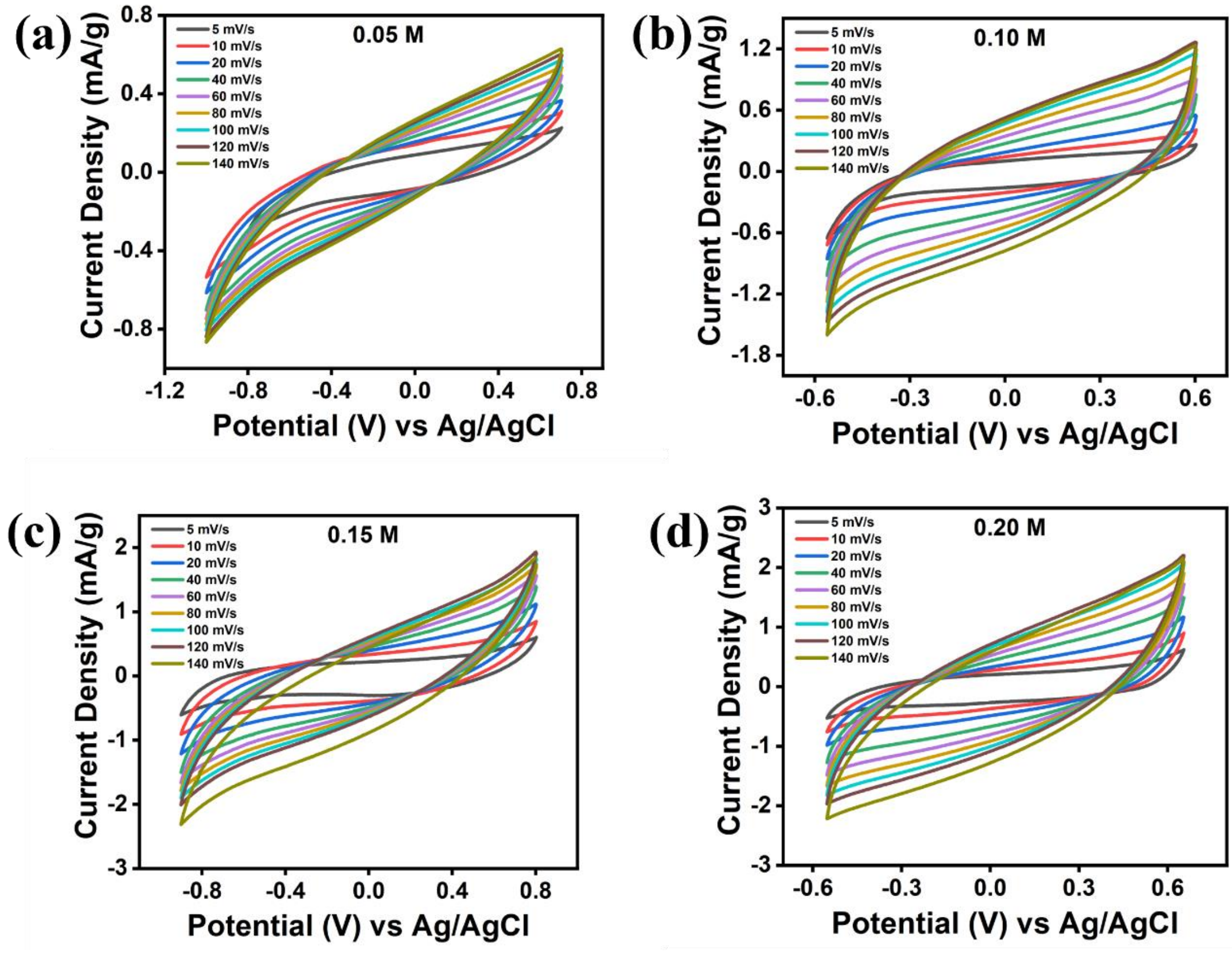

**Figure S7.** Cyclic voltammograms at 5–140 mV s⁻¹ for PAA binder at all four LiTFSI concentrations.

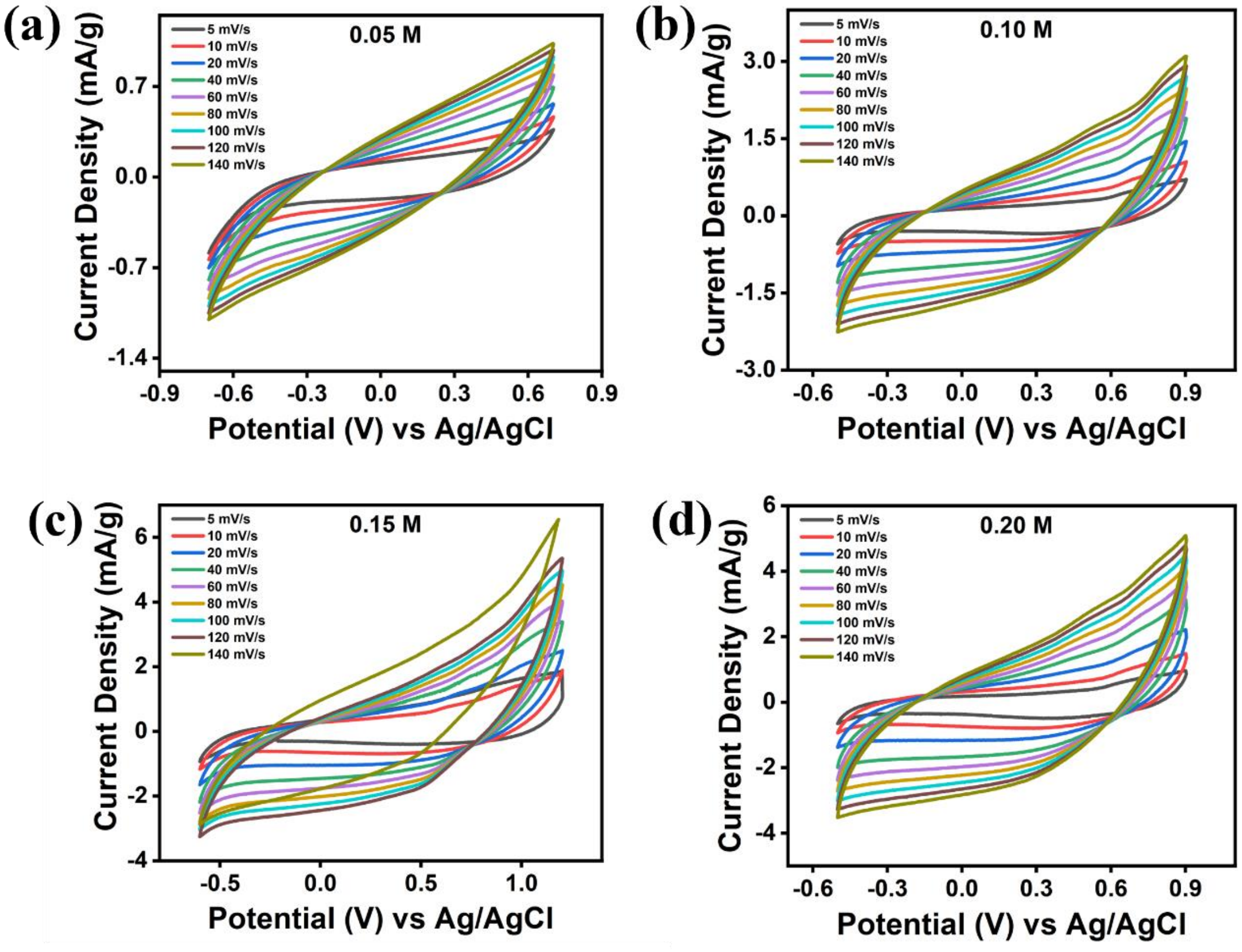

**Figure S8.** Cyclic voltammograms at 5–140 mV s⁻¹ for PEDOT:PSS binder at all four LiTFSI concentrations.

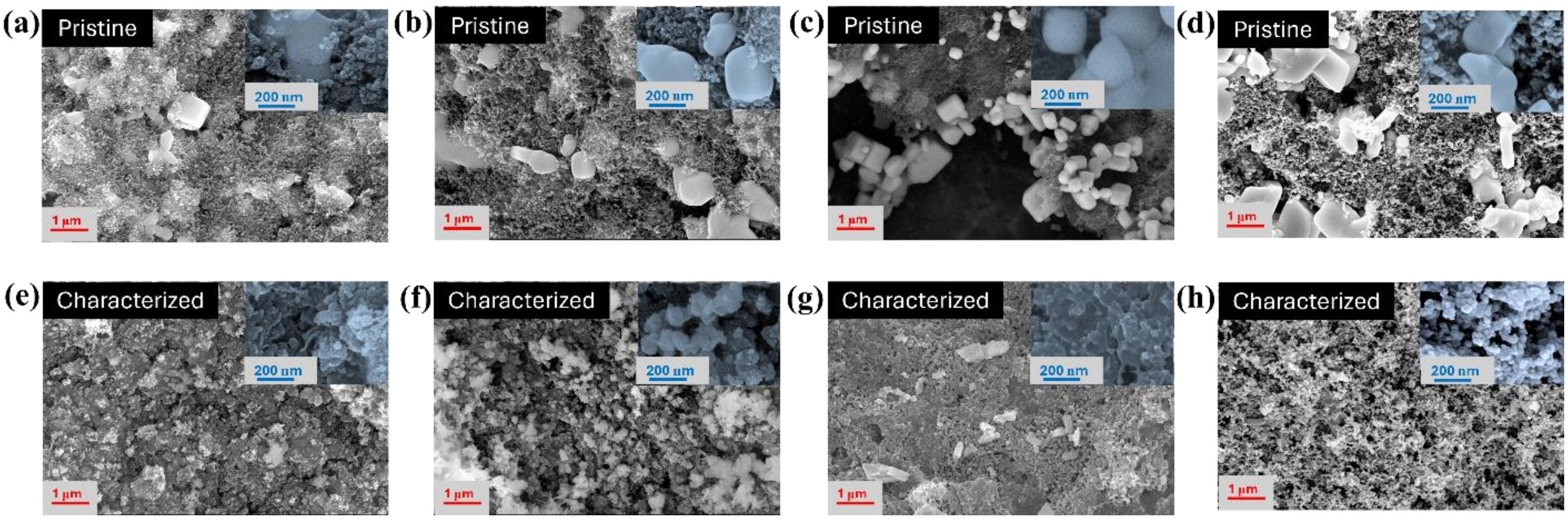

**Figure S9.** Field-emission scanning electron micrographs of all four electrodes before and after electrochemical characterization.

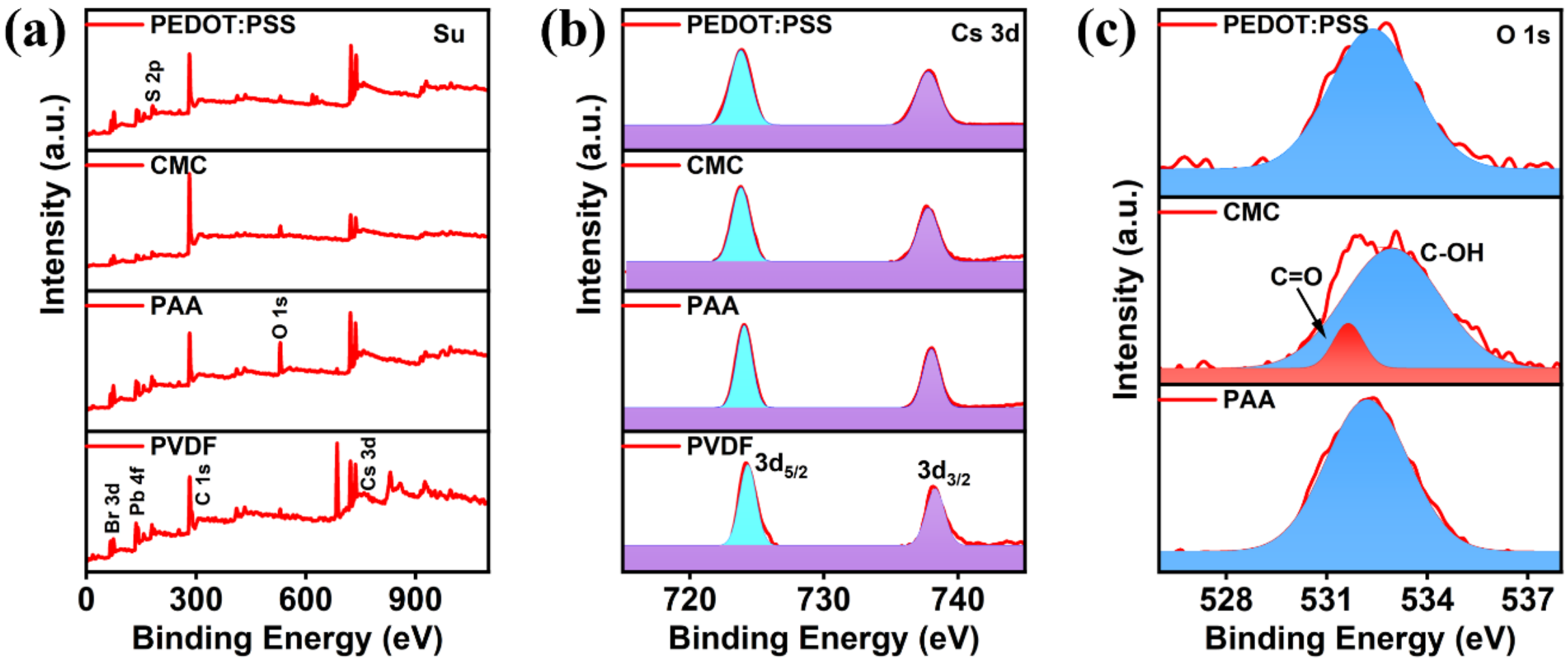

**Figure S10.** Supporting X-ray photoelectron spectra: survey scans, Cs 3d for all four pristine electrodes, O 1s for PAA, CMC and PEDOT:PSS.